\documentclass[12pt,letterpaper]{article}
\usepackage[usenames, dvipsnames]{xcolor}
\usepackage{jcapmod}

\usepackage{tocloft}
\usepackage[]{todonotes}
\usepackage{verbatim}
\usepackage{mathrsfs}
\usepackage{cleveref}
\usepackage[shortlabels]{enumitem}
\usepackage{pgf}
\usepackage{tikz-cd}
\usetikzlibrary{shapes,arrows}
\usetikzlibrary{calc}
\usepackage{pgfplots}
\pgfplotsset{compat=1.12}
\usepackage{tikz-3dplot}

\usepackage{amsthm}

\usepackage{tikz}
\usepackage{setspace,caption}
\usepackage{lipsum}
\usetikzlibrary{matrix,arrows,calc}
\makeatletter
\newsavebox\myboxA
\newsavebox\myboxB
\newlength\mylenA

\definecolor{cornellRed}{HTML}{B31B1B}
\definecolor{cornellBlue}{HTML}{0068AC}
\definecolor{cornellGreen}{HTML}{6EB43F}

\usepackage{bbm} 					
\usepackage{slashed} 				
\usepackage{graphicx}				
\usepackage{subcaption}			
\usepackage{psfrag}				
\usepackage{tensor}				
\usepackage{fouridx}				
\usepackage{bm}					
\usepackage{mdframed}				
\usepackage{multirow}				
\usepackage{soul}					
\usepackage{bbold}				
\usepackage{multicol}				
\usepackage{tikz-cd}
\usepackage{rotating}

\usetikzlibrary{arrows}

\tikzset{
commutative diagrams/.cd,
arrow style=tikz,
diagrams={>=latex}}

\usepackage{amsmath}
\usepackage{amssymb}
\usepackage{amsfonts}
\usepackage{mathtools}

\usepackage{feynmf}
\usepackage{marvosym}

\usepackage{import}

\newcommand*\xoverline[2][0.75]{%
    \sbox{\myboxA}{$\m@th#2$}%
    \setbox\myboxB\null
    \ht\myboxB=\ht\myboxA%
    \dp\myboxB=\dp\myboxA%
    \wd\myboxB=#1\wd\myboxA
    \sbox\myboxB{$\m@th\overline{\copy\myboxB}$}
    \setlength\mylenA{\the\wd\myboxA}
    \addtolength\mylenA{-\the\wd\myboxB}%
    \ifdim\wd\myboxB<\wd\myboxA%
       \rlap{\hskip 0.5\mylenA\usebox\myboxB}{\usebox\myboxA}%
    \else
        \hskip -0.5\mylenA\rlap{\usebox\myboxA}{\hskip 0.5\mylenA\usebox\myboxB}%
    \fi}
\makeatother

\newcommand{\im}{\,\mathrm{Im}\,}
\newcommand{\re}{\,\mathrm{Re}\,}

\definecolor{cobalt}{RGB}{44, 98, 120}
\definecolor{celadon}{rgb}{0.67, 0.88, 0.69}
\definecolor{dm}{cmyk}{.20, 0, .30, 0}
\definecolor{burgundy}{rgb}{0.5, 0.0, 0.13}
\definecolor{plotBlue}{RGB}{94, 130, 181}

\hypersetup{
  colorlinks,
  citecolor=Violet,
  linkcolor=cobalt,
  urlcolor=Blue}

\DeclareSymbolFontAlphabet{\mathbb}{AMSb}

\newif\iffastcompile

\fastcompilefalse

\iffastcompile
\newcommand{\mk}[1]{}
\else
\newcommand{\mk}[1]{\todo[color=burgundy!30, size=\scriptsize, bordercolor=burgundy!30]{MK: #1}}
\fi

\ProvideTextCommandDefault{\Dbar}{%
\leavevmode\lower.5ex\rlap{\hskip-.07em\accent"16}D%
}

\usepackage{environ}
\usepackage{changepage}

\begin{document}
	\newcommand{\main}{.}
\begin{titlepage}

\setcounter{page}{1} \baselineskip=15.5pt \thispagestyle{empty}
\setcounter{tocdepth}{2}
	{\hfill \small MIT-CTP/5531}
\bigskip\

\vspace{1cm}
\begin{center}
{\large \bfseries On string one-loop correction to the  Einstein-Hilbert term}\\~\\

{\large \bfseries and}\\~\\

{\large \bfseries its implications on the K{\"a}hler potential. }
\end{center}

\vspace{0.55cm}

\begin{center}
\scalebox{0.95}[0.95]{{\fontsize{14}{30}\selectfont Manki Kim$^{a}$\vspace{0.25cm}}}

\end{center}

\begin{center}

\vspace{0.15 cm}
{\fontsize{11}{30}
\textsl{$^{a}$Center for Theoretical Physics, Department of Physics, Massachusetts Institute of Technology, Cambridge, MA 02139}}\\
\vspace{0.25cm}

\vskip .5cm
\end{center}

\vspace{0.8cm}
\noindent
To compute the string one-loop correction to the K{\"a}hler potential of moduli fields of string compactifications in Einstein-frame, one must compute: the string one-loop correction to the Einstein-Hilbert action, the string one-loop correction to the moduli kinetic terms, the string one-loop correction to the definition of the holomorphic coordinates. In this note, in the small warping limit, we compute the string one-loop correction to the Einstein-Hilbert action of type II string theory compactified on orientifolds of Calabi-Yau threefolds. We find that the one-loop correction is determined by the new supersymmetric index studied by Cecotti, Fendley, Intriligator, and Vafa and the Witten index. As a simple application, we apply our results to estimate the size of the one-loop corrections around a conifold point in the K{\"a}hler moduli space.

\vspace{1.1cm}

\vspace{3.1cm}

\noindent\today

\end{titlepage}
\tableofcontents\newpage

\section{Introduction}
One of the most pressing questions in string phenomenology is to exhibit cosmological solutions that share many features of our universe. As an example, as our universe appears to be well approximated by quasi de Sitter solution, it is extremely important to understand if string theory can populate meta-stable de Sitter vacua with small cosmological constant and large decay time.

We face an imminent challenge. The more realistic we want our solutions to be, the fewer theoretical tools are at our disposal to help us handle string theory directly, even perturbatively. This is mainly because more realistic solutions are expected to have no supersymmetry and the worldsheet theory in such cases is not expected to be weakly coupled. Because of our incomplete understanding of strongly coupled worldsheet theory and the non-perturbative formulation of string theory, it is very difficult to envision a scenario in which one finds a set of isolated cosmological solutions of string theory by the means of finding appropriate worldsheet CFTs. 

However, one shouldn't be completely discouraged by our lack of understanding of the nonperturbative and strongly coupled phase of string theory. Because finding vacuum solutions of string theory is inherently a low-energy problem, one might still hope to make progress by understanding the low-energy effective theory of string compactifications. Progress along this line of work will involve judiciously deriving the low-energy effective action of string compactifications and finding isolated vacua thereof. 

Nevertheless, one could reasonably complain that the usage of the low-energy effective theory does not seem to circumvent the problem by much. The nature of the problem is that to have a non-trivial vacuum solution, many terms in the effective potential shall compete, which indicates that the underlying string theory is presumably strongly coupled. In light of the fact that it is extremely difficult to compute anything beyond string tree level, naively, it may seem practically impossible to derive the low-energy action with an arbitrary precision to find isolated vacua of string theory. This problem is so-called the Dine-Seiberg problem \cite{Dine:1985he}.

In 2003, by Kachru, Kallosh, Linde, and Trivedi (KKLT), a three-step recipe to overcome the Dine-Seiberg problem to find de Sitter vacua of string theory is proposed \cite{Kachru:2003aw}.\footnote{For a very closely related proposal, see \cite{Balasubramanian:2005zx}.} The KKLT scenario proceeds in type IIB compactifications on O3/O7 orientifolds of Calabi-Yau threefolds, that yields 4d $\mathcal{N}=1$ supersymmetric theory at low energy, as follows. First, stabilize complex structure moduli and the axio-dilaton by the Gukov-Vafa-Witten (GVW) flux superpotential \cite{Gukov:1999ya,Giddings:2001yu}
\begin{equation}
W_{GVW}=\int G\wedge \Omega\,,
\end{equation}
at a point in the moduli space such that vacuum expectation value $W_0:=\langle W_{GVW}\rangle $ is exponentially small.\footnote{For recent progress on engineering exponentially small flux superpotential, see \cite{Demirtas:2019sip,Demirtas:2020ffz,Cicoli:2022vny,Carta:2022oex,Grimm:2021ckh,Bastian:2021hpc,Broeckel:2021uty, Alvarez-Garcia:2020pxd}.} Second, stabilize K{\"a}hler moduli by balancing the non-perturbative superpotential $\mathcal{A} e^{-a T}$ \cite{Witten:1996bn} against $W_0.$ At this step, exponentially small value of $W_0$ guarantees that $T$ is stabilized at large Einstein-frame volume. As a result of the first two steps, one arrives at 4d $\mathcal{N}=1$ supersymmetric vacua with exponential scale separation. As the last step, one can break supersymmetry by placing anti-D3-branes at the tip of the Klebanov-Strassler throat to engineer low-energy supersymmetry breaking and attain de Sitter solutions \cite{Klebanov:2000hb,Kachru:2002gs}.\footnote{For recent studies on stability issues of strongly warped regions and anti-D3-brane supersymmetry breaking, see for example \cite{Bena:2018fqc,Randall:2019ent,Dudas:2019pls,Gao:2020xqh,Carta:2021lqg,Lust:2022xoq,Hebecker:2022zme,Schreyer:2022len}.} 

One can ask, how does the KKLT proposal tame the Dine-Seiberg problem? The crucial insight is that one can in principle compute the superpotential with arbitrary precisions, and by doing so one can stabilize moduli at weak coupling and large Einstein-frame volume even if our knowledge on the K{\"a}hler potential is limited.\footnote{For recent progress on explicit constructions of KKLT like 4d $\mathcal{N}=1$ vacua, see \cite{Demirtas:2021nlu}.} Because in type IIB compactifications on O3/O7 orientifolds, the good holomorphic coordinate for K{\"a}hler moduli is Einstein-frame divisor volume, even if string-frame cycle volumes are order $\mathcal{O}(1),$ as long as string coupling is small, overall Calabi-Yau volume is large, worldsheet instanton corrections to the K{\"a}hler potential are suppressed, and Einstein-frame divisor volume is large, one can expect that the string loop correction to the K{\"a}hler  potential is suppressed by the order of
\begin{equation}
\mathcal{O}\left(\frac{g_s^2}{\mathcal{V}}\right)\,,
\end{equation}
where $\mathcal{V}$ is string-frame Calabi-Yau volume.\footnote{For a more detailed analysis of the order estimate of the loop corrections to the K{\"a}hler potential, see section 4 of \cite{Demirtas:2021nlu}.} But, there can be an unexpectedly large one-loop correction to the K{\"a}hler potential due to a numerical coincidence. To put our understanding of vacuum solutions of string theory on more solid footing, we should therefore actually compute the one-loop correction to the K{\"a}hler potential.

In the literature, results on the loop corrections to the K{\"a}hler potential in $\mathcal{N}=1$ compactifications are scarce \cite{Berg:2005ja,Berg:2007wt,Cicoli:2007xp,Berg:2011ij,Berg:2014ama,Haack:2015pbv,Haack:2018ufg,Gao:2022uop}. One of the main reasons why it has been difficult to compute the loop corrections to the K{\"a}hler potential, unlike the computation of the superpotential, is because the K{\"a}hler potential isn't a holomorphic object. Due to the lack of holomorphicity, it is not clear whether dualities or supersymmetry will help us compute the K{\"a}hler potential in a rather simple manner. Therefore, it is extremely important to develop tools to explicitly and directly compute the loop corrections to the K{\"a}hler potential.

We would like to highlight the heroic computations carried out initially by Berg, Haack, and Kors (BHK) \cite{Berg:2005ja}, and the subsequent follow up papers by Berg, Haack, and their collaborators \cite{Berg:2007wt,Berg:2011ij,Berg:2014ama,Haack:2015pbv,Haack:2018ufg}. In these papers, to determine string one-loop corrected K{\"a}hler metric in Einstein-frame, very impressive computations were carried out in toroidal orientifold compactifications to determine: the string one-loop correction to the Einstein-Hilbert (EH) term in string-frame, the string one-loop correction to the moduli kinetic terms in string-frame, and the string one-loop correction to the definition of holomorphic coordinates. It is intuitive to see why the string one-loop corrections to the moduli kinetic terms and the definition of holomorphic coordinates need to be computed. The necessity of the knowledge on the string one-loop corrected Einstein-Hilbert term is due to the fact that changing from string-frame to Einstein-frame will force the one-loop corrected K{\"a}hler potential to depend on the string one-loop corrected EH term. 

In this note, we will take a first step at generalizing the works done by \cite{Berg:2005ja,Berg:2007wt,Berg:2011ij,Berg:2014ama,Haack:2015pbv,Haack:2018ufg} to genuine type II compactifications on orientifolds of Calabi-Yau threefolds. To do so, we will compute the two graviton scattering amplitudes with several restrictions we impose to simplify the analysis. In this work, we shall perturb around the Calabi-Yau background by placing spacetime filling D-branes and O-planes on the Calabi-Yau background. Additionally, we shall assume that the backreaction from the spacetime filling D-branes and O-planes is negligible. Phrased differently, we shall take the worldsheet CFT to be a direct sum of the free field CFT for the non-compact directions and the Calabi-Yau CFT for the compact directions. D-branes and O-planes shall be described as boundary states of the aforementioned CFT. This approximation is well warranted for weak warping, where the coupling between the non-compact directions and the compact directions is negligible. Furthermore, we shall cancel the Ramond-Ramond tadpole by spacetime filling D-branes. This will allow us to use the RNS formalism for the worldsheet theory.

One immediate concern follows. It is a well-known lore in string theory that every two-point amplitude of massless fields vanishes once the momentum conservation condition and the transversality condition are imposed. The graviton two-point functions, that we compute in this draft, are no exceptions from this well-known statement. This phenomenon is not unique to string theory. Even in quantum field theory, to compute the wave function renormalization of a field, one oftentimes relaxes the momentum conservation condition for two-point functions. Therefore, to extract the correction to the EH action from the graviton two point amplitudes, we have to invoke a prescription for relaxing the momentum conservation condition. Prescriptions that are reasonable for quantum field theories have no right to be reasonable for string scattering amplitudes. Nonetheless, inspired by quantum field theory, and encouraged by successes in simpler settings \cite{Antoniadis:1996vw,Haack:2015pbv}, we will make use of the prescription of relaxing the momentum conservation condition, that we explain in \S\ref{sec:graviton scattering at one loop}, to extract the string one-loop correction to the Einstein-Hilbert action. It would be extremely important to cross-check the results obtained by using the prescription via unambiguous string amplitudes, such as four graviton scattering amplitudes. As a non-trivial cross-check of the prescription, we will find the perfect match between the torus amplitude we compute in \S\ref{sec:torus} with by now a very well-established result in the literature \cite{Kiritsis:1994ta,Antoniadis:1996vw,Antoniadis:1997eg,Antoniadis:2002tr, Green:1997di,Green:1997as,Kiritsis:1997em,Russo:1997mk,Becker:2002nn,Liu:2022bfg}.

In this work, by explicit computations, we will claim that the string one-loop correction to the EH action $\delta E$ is determined by the new supersymmetric index \emph{new supersymmetric index} studied by Cecotti, Fendley, Intriligator, and Vafa \cite{Cecotti:1992qh}
\begin{equation}
\text{Tr}_R \left((-1)^{F-\frac{3}{2}}F q^{L_0-\frac{3}{8}}\right)\,,
\end{equation}
and the Witten index of the internal CFT. For example, we will find that the annulus contribution reads
\begin{equation}\label{eqn:main00}
\boxed{\delta E_A=\frac{1}{2^9\pi^2}\int_0^\infty \frac{dt}{2t^2} \Bigl[\text{Tr}_R \left((-1)^{F-\frac{3}{2}}F q^{L_0-\frac{3}{8}}\right)^{open}_{int}+\frac{3}{2} \left(n^+_A-n^-_A\right)\Bigr]\,,}
\end{equation}
where we parametrize the one-loop corrections to the EH term as
\begin{equation}
\frac{1}{2\kappa_4^2}\int d^4 x\sqrt{-g}\, \delta E\, R\,.
\end{equation}
It is important to note that despite the name, the new supersymmetric index is not a number but a complicated function of moduli. But, the new supersymmetric index is special because it only depends on the F-term of the internal theory, it might be possible to compute it in excruciating detail in explicit models. Note that the trace is taken over the internal CFT. This claim is surprising because the new supersymmetric index is expected to show up in protected quantities, whereas the EH action are not protected in the supergravity. In \S\ref{sec:discussion}, we will provide heuristic explanations on this unexpected result.

The rest of the paper is organized as follows. In \S\ref{sec:Kahler potential}, we review how to translate the string one-loop corrected effective action in string-frame to the string one-loop corrected effective action in Einstein-frame. We will study in the same section that the one-loop corrected K{\"a}hler potential depends on the string one-loop correction to the EH action. In \S\ref{sec:graviton scattering at one loop}, we compute graviton-graviton scattering at string one-loop to compute the string one-loop correction to the EH action. In \S\ref{sec:small cycles}, we apply our results to estimate the one-loop correction to the EH action and the Kahler potential when one is approaching a conifold singularity, which is a finite distance singularity. In \S\ref{sec:discussion}, we conclude and discuss possible future directions. In \S\ref{app:jacobi theta}, we summarize useful identities involving the Jacobi theta functions. In \S\ref{app:N=2 superconformal algebra}, we review the extended $\mathcal{N}=2$ superconformal algebra and its representation theory. In \S\ref{app:Greens function}, we summarize useful identities involving Green's functions. In \S\ref{app:two point functions}, we compute various two-point functions.

\section{One-loop corrections to the K{\"a}hler potential}\label{sec:Kahler potential}
To translate string scattering amplitudes into supergravity actions, one must compare scattering amplitudes of string theory with scattering amplitudes computed from supergravity action at low energy. Quite naturally, this comparison is the most straightforward in string-frame, as the name speaks for itself. On the other hand, in the supergravity description, varying the Planck constant is very unnatural, and therefore one oftentimes works in Einstein frame, in which the Planck scale is set to a constant. This difference in the use of frame results in a non-trivial dictionary between string scattering amplitudes at one-loop and the string-loop corrections to the supergravity action in Einstein frame. In this section, following \cite{Berg:2014ama}, we shall review how one-loop amplitudes of string theory can be repackaged in terms of the one-loop correction to the K\"{a}hler potential of moduli fields in \emph{Einstein}-frame. To simplify the discussion, we will focus on one modulus $\phi.$ 

Let us start with string tree-level supergravity action in string-frame with most two-derivatives 
\begin{equation}\label{eqn:treelevel action string frame}
S^{(0)}_{st}\supset \frac{1}{\kappa_4^2}\int d^4 x\sqrt{-g}e^{-2\Phi_4}\left[\frac{1}{2} R -G^{(0)}_{\varphi\varphi}\partial_\mu \varphi^{(0)}\partial^{\mu}\varphi^{(0)}\right]\,,
\end{equation}
where $\varphi,$ we denote a saxion field which is the imaginary part of the chiral field $\phi.$ If we are in a geometric phase, in large volume limit, we have a relation
\begin{equation}
e^{-2\Phi_{4}}:=e^{-2\Phi_{10}}\mathcal{V}\,,
\end{equation}
where $\mathcal{V}$ is the volume of the compactification manifold in string unit, and $\Phi_{10}$ is the 10-dimensional dilaton which is identified with string coupling as $g_c=e^{\Phi_{10}}.$ This tree-level action can be computed by comparing the low-energy expansion of string scattering amplitudes at the sphere level to scattering amplitudes of supergravity fields. Let us now transform the supergravity action in string-frame \eqref{eqn:treelevel action string frame} into the supergravity action in Einstein-frame by rescaling the metric
\begin{equation}
g= e^{2\Phi_4} \tilde{g}\,.
\end{equation}
Then, we have
\begin{equation}
\sqrt{-g} = e^{4 \Phi_4}\sqrt{-\tilde{g}}\,,
\end{equation}
and
\begin{equation}
R= e^{-2\Phi_4} \tilde{R}-6e^{-2\Phi_4}\nabla \Phi_4-6e^{-2\Phi_4}\partial_\mu \Phi_4 \partial^\mu \Phi_4\,.
\end{equation}
As a result, we find
\begin{equation}
S_{E}^{(0)}\supset \frac{1}{\kappa_4^2}\int d^4 x \sqrt{-\tilde{g}} \left[\frac{1}{2}\tilde{R} -G_{\varphi\varphi}^{(0)}\partial_\mu \varphi^{(0)}\partial^\mu\varphi^{(0)}-3\partial_\mu \Phi_4\partial^\mu \Phi_4\right]\,.
\end{equation}
Therefore the kinetic term of the modulus in Einstein-frame reads
\begin{equation}
-\frac{1}{\kappa_4^2}\int d^4x \sqrt{-\tilde{g}} \,\left(G_{\varphi\bar{\varphi}}^{(0)}+3(\partial_{\varphi}\Phi_4)^2\right)\partial_\mu \varphi^{(0)}\partial^\mu\varphi^{(0)}\,.
\end{equation}
As the metric of the modulus must be K{\"a}hler, we conclude that the tree-level K{\"a}hler metric of $\phi$ is given as
\begin{equation}
K_{\phi\bar{\phi}}^{(0)}:=G_{\varphi\varphi}^{(0)}+3(\partial_\varphi \Phi_4)^2\,.
\end{equation}

Let us now suppose that one computed string scattering amplitudes at string one-loop to obtain supergravity action at string one-loop in string-frame
\begin{equation}
S^{(0)}_{st}+S^{(1)}_{st}\supset \frac{1}{\kappa_4^2}\int d^4 x\sqrt{-g} \left[ \frac{1}{2} \left(e^{-2\Phi_4}+\delta E\right) R+(e^{-2\Phi_4}G_{\varphi\varphi}^{(0)} +G_{\varphi\varphi}^{(1)})\partial_\mu \varphi^{(0)}\partial^\mu \varphi^{(0)}\right]\,.
\end{equation}
Note that the definition of moduli fields also gets corrected at one-loop \cite{Berg:2014ama,Haack:2018ufg}
\begin{equation}
\phi=\phi^{(0)}+\phi^{(1)}\,.
\end{equation}

We shall now find the effective action in Einstein frame. Let us define
\begin{equation}
e^{-2\tilde{\Phi}_4}:= e^{-2\Phi_4}+\delta E\,.
\end{equation}
Let us rescale the metric
\begin{equation}
g=e^{2\tilde{\Phi}_4}\tilde{g}\,,
\end{equation}
to find
\begin{align}
S_{E}^{(0)}+S_E^{(1)}\supset &\frac{1}{\kappa_4^2}\int d^4 x \sqrt{-\tilde{g}} \biggl[\frac{1}{2}\tilde{R} -\Bigl((1-\delta E e^{2\Phi_4})G_{\varphi\varphi}^{(0)}+e^{2\Phi_4}G_{\varphi\varphi}^{(1)}\Bigr)\partial_\mu \varphi^{(0)}\partial^\mu\varphi^{(0)}\nonumber\\
&\qquad\qquad\quad-3\Bigl((1-2\delta E e^{2\Phi_4}) \partial_\mu\Phi_4\partial^\mu \Phi_4 -e^{2\Phi_4}\partial_\mu\Phi_4 \partial^\mu (\delta E) \Bigr)\biggr]\,.
\end{align}
By matching the above action to the one-loop corrected kinetic term of the modulus $\phi$
\begin{equation}
-\frac{1}{\kappa_4^2}\int d^4x \sqrt{-\tilde{g}}\left(K_{\phi\bar{\phi}}^{(0)}+e^{2\Phi_4}K_{\phi\bar{\phi}}^{(1)} \right)\partial_\mu \phi\partial^\mu\bar{\phi}\,,
\end{equation}
we find
\begin{equation}
K_{\phi\bar{\phi}}^{(1)}=G_{\varphi\varphi}^{(1)}-6\left(\frac{\partial\Phi_4}{\partial \phi^{(0)}}\right)^2\delta E -3\frac{\partial\Phi_4}{\partial\phi^{(0)}}\frac{\partial (\delta E)}{\partial \phi^{(0)}}-G_{\varphi\varphi}^{(0)}\delta E+\frac{1}{2\phi^3}\phi^{(1)}-\frac{1}{2\phi^2}\frac{\partial\phi^{(1)}}{\partial \phi}\,.
\end{equation}
Therefore, we conclude that to completely determine the one-loop corrected K{\"a}hler potential for moduli fields in Einstein frame, computing the one-loop correction to the Einstein-Hilbert action is necessary. 

\section{Two graviton scattering at one-loop}\label{sec:graviton scattering at one loop}
In this section, we will determine the string one-loop corrections to the Einstein-Hilbert action by computing the two-graviton scattering amplitudes at string one-loop in type II string theory compactified on an orientifold of a Calabi-Yau threefold. 

\subsection{Strategy}
In this section, we shall explain the strategy to compute the string one-loop corrections to the Einstein-Hilbert action.

A prerequisite to the computation of string scattering amplitudes at the string one-loop, is to first construct a consistent worldsheet CFT at string tree level. In type II compactifications on Calabi-Yau manifolds, it is well known that the exact worldsheet CFT is given as a direct sum of the free field CFT, that describes the non-compact directions, and the Calabi-Yau CFT.  But, in orientifold compactifications, such an exact description of the worldsheet CFT is not known. The main complication is that spacetime filling D-branes and O-planes backreact on the target space, and as a result a non-trivial Ramond-Ramond profile and a non-trivial mixing between the non-compact directions and the compact directions are generated.
 
Even if the Ramond-Ramond tadpole is canceled by spacetime filling D-branes, such a backreaction can be significant unless the Ramond-Ramond charge is canceled locally, meaning $2^{p-5}$ Dp-branes lie on top of an Op-plane. But, it should be noted that in generic Calabi-Yau orientifold compactifications, the Dp-brane tadpole can come from not only spacetime filling Op-planes, but also from the curvature correction terms to higher dimensional spacetime filling D-branes and O-planes. For example, it is well known that the curvature correction to a D7-brane wrapped on a divisor $D$ can induce the following D3-brane tadpole \cite{Denef:2008wq}
\begin{equation}
\frac{\chi(D)}{24}\,.
\end{equation}

In light of this difficult problem,  we shall invoke an approximation scheme. Suppose that there is a net charge $N_{Dp}$ of a wrapped Dp-brane and the Einstein-frame transverse volume to such a Dp-brane is $\mathcal{V}_\perp.$ Then, we shall require that 
\begin{equation}
\frac{N_{Dp}}{\mathcal{V}_\perp}\ll1\,,
\end{equation}
such that the backreaction from the spacetime filling D-branes and O-planes is negligible. For example, in type IIB compactifications on O3/O7 orientifolds of Calabi-Yau threefolds, we shall require that given the number of spacetime filling D3-branes, $N_{D3},$ and the Einstein-frame volume of the Calabi-Yau manifold, $\mathcal{V},$ the following inequality holds
\begin{equation}
\frac{N_{D3}}{\mathcal{V}}\ll 1\,,
\end{equation}
and the D7-brane charge is locally canceled. We will call this approximation the small warping approximation or the small warping limit.

In the small warping approximation, we can now treat the worldsheet CFT as a direct sum of the free field CFT and the Calabi-Yau CFT, as in the Calabi-Yau compactifications. Spacetime filling D-branes and O-planes will then be described as the boundary states of the aforementioned CFT even at string one-loop. With this understanding, we shall finally spell out the details of the worldsheet CFT.

The worldsheet theory contains the $b,~c,~\beta,~\gamma$ ghost system, the fields $\xi,~\eta,~\phi,$ obtained by bosonizing the $\beta,~\gamma$ system, the matter superconformal field theory. The matter part of the CFT is decomposed into $(1,1)$ supersymmetric free field CFT with central charge $(c,\bar{c})=(6,6)$ that describes the four non-compact space-time, and a strongly interacting $(2,2)$ supersymmetric SCFT with central charge $(c,\bar{c})=(9,9)$ describing the Calabi-Yau manifold. By $X^\mu$ and $\psi^\mu,$ we denote the matter fields and their superpartner that correspond to coordinates and the tangent bundle of the non-compact spacetime, respectively. To obtain $\mathcal{N}=1$ effective theory in target spacetime, we need to reduce the amount of supersymmetry by half. To reduce the target space supersymmetry by half, we will perform an orientifolding $\Omega$ and project out degrees of freedom. To simplify the computation, we will assume that the RR tadpole is canceled by spacetime filling D-branes.

To specify the orientifolding, we will have to study the orientifold action on worldsheet fields and coordinates. As the name stands, the orientifold action $\Omega$ reverses the orientation of the worldsheet by mapping the worldsheet coordinate $z$ to $-\bar{z}.$\footnote{For the action of the orientifold on genus-1 surfaces, see \S\ref{app:two point functions}.} We note that the fields in the free CFT transform under the orientifolding as follows
\begin{align}
&\partial X^\mu (z)\mapsto -\partial X^\mu (-\bar{z})\,,~ c(z)\mapsto -\bar{c}(-\bar{z})\,,~ b(z)\mapsto\bar{b}(-\bar{z})\,, ~ e^{-\phi}\psi^\mu(z)\mapsto -e^{-\bar{\phi}}\bar{\psi}^\mu(-\bar{z})\,,\\
& \partial\xi(z)\mapsto -\bar{\partial}\bar{\xi}(-\bar{z})\,,~ \eta(z)\mapsto -\bar{\eta}(-\bar{z})\,,~ e^{-2\phi}(z)\mapsto e^{-2\bar{\phi}}(-\bar{z})\,.\nonumber
\end{align}
As the details of the orientifold action on the internal CFT are not very important for our discussion, we will omit such details. For the detail of the orientifold action, see for example \cite{Alexandrov:2022mmy}.

Let us study the orientifold action on the graviton vertex operators. We write the graviton vertex operators as 
\begin{equation}
V_g^{(0,0)}(z,\bar{z})=-\frac{2g_c}{\alpha'}\epsilon_{\mu\nu} \left(i\partial X^\mu +\frac{\alpha'}{2}p \cdot \psi \psi^\mu\right)\left(i\bar{\partial}X^\nu + \frac{\alpha'}{2} p\cdot \bar{\psi}\bar{\psi}^\nu\right) e^{i p\cdot X}\,,
\end{equation}
\begin{equation}
V_g^{(0,-1)}=i g_c\sqrt{\frac{2}{\alpha'}}\epsilon_{\mu\nu} \left(i\partial X^\mu +\frac{\alpha'}{2}p \cdot \psi \psi^\mu\right) e^{-\bar{\phi}}\bar{\psi}^{\nu}e^{ip\cdot X}\,,
\end{equation}
\begin{equation}
V_g^{(-1,0)}=i g_c\sqrt{\frac{2}{\alpha'}}\epsilon_{\mu\nu} e^{-\phi} \psi^\mu\left(i\bar{\partial}X^\nu + \frac{\alpha'}{2} p\cdot \bar{\psi}\bar{\psi}^\nu\right) e^{ip\cdot X}\,,
\end{equation}
\begin{equation}
V_g^{(-1,-1)}(z,\bar{z})=g_c \epsilon_{\mu\nu}e^{-\phi} \psi^\mu e^{-\bar{\phi}}\bar{\psi}^\nu e^{ip \cdot X}\,,
\end{equation}
where $\epsilon_{\mu\nu}$ is the polarization tensor with $\epsilon_{\mu\nu}\epsilon^{\mu\nu}=1.$ Note that the graviton vertex operators are constructed purely from the matter fields of the free part of the CFT. We find that the orientifold action $\Omega$ maps the graviton vertex operator with a picture number $(p,q)$ to itself
\begin{equation}
\Omega:V_g^{(p,q)}\mapsto V_g^{(p,q)}\,.
\end{equation}
One very important remark is in order. One might be tempted to conclude that sign of the orientifold action on the $SL(2,\Bbb{C})$ invariant vacuum $|0\rangle$ must be chosen to be even to keep the graviton in the spectrum. But, this conclusion is too quick. It should be noted that when a vertex operator is inserted and its position is fixed by the $SL(2,\Bbb{C})$ invariance, one must also insert an appropriate number of $c, \bar{c}$ ghosts, $c\bar{c}$ for the graviton vertex operator. On the other hand, if the position of a vertex operator is not fixed, one must integrate over the possible positions of the vertex operator by including the measure factor $\int d^2z,$ where $z$ is the position of the vertex operator. Because $c\bar{c}$ and $\int d^2 z$ are odd under the orientifold action $\Omega,$ the graviton state prepared by inserting the graviton vertex operator at the origin is therefore also odd, concerning the orientifold action on $|0\rangle,$ under the orientifold action. This fixes the sign of the orientifold action of the $SL(2,\Bbb{C})$ invariant vacuum to be odd.

We shall now study how to make a comparison between the one-loop scattering amplitude and the string one-loop correction to the Einstein-Hilbert action. Let us start by writing the one-loop correction to the Einstein-Hilbert action
\begin{equation}
S^{(1)}_{st}\supset\frac{1}{2\kappa_4^2}\int d^4 x \sqrt{-g}\, \delta E\,R  \,,
\end{equation}
where $\delta E$, in general, depends on moduli. Let us now expand
\begin{equation}
g_{\mu\nu}=\eta_{\mu\nu}+h_{\mu\nu}\,,
\end{equation}
where $\eta_{\mu\nu}$ is the Minkowski metric, then we obtain
\begin{align}
S^{(1)}_{st}\supset& \frac{1}{4\kappa_4^2}\int d^4x \delta E\, \left[ (\partial_\mu h^{\mu\nu})(\partial_\nu h)-(\partial_\mu h^{\rho\sigma}) (\partial_\rho h^{\mu}_{~\sigma})+\frac{1}{2} \eta^{\mu\nu}(\partial_\mu h^{\rho\sigma})(\partial_\nu h_{\rho\sigma})\right.\nonumber\\
&\left.\qquad\qquad\qquad\qquad-\frac{1}{2}\eta^{\mu\nu}(\partial_\mu h)(\partial_\nu h)\right]+\mathcal{O}(h^3)\,.
\end{align}

We aim to match this action by comparing it to the one-loop scattering amplitudes. Let us consider the graviton two-point function computed on the genus 1 Riemann surfaces with or without boundaries
\begin{equation}\label{eqn:graviton two point}
Z=\sum_{R}\langle\langle V_g(p_1,\epsilon_1)V_g(p_2,\epsilon_2)\rangle\rangle_{R}\,,
\end{equation}
where $R$ runs over the torus, annuli, M{\"o}bius strip, and the Klein bottle, $p_i$ are graviton momentum, and $\epsilon_i$ is the polarization tensors for the gravitons. Formally, the two-point function \eqref{eqn:graviton two point} vanishes once the on-shell conditions and the momentum conservation condition
\begin{equation}
p_1^2=p_2^2=p_1+p_2=p_1\cdot p_2=p_{1\mu}\epsilon^{\mu\nu}_1=p_{2\mu}\epsilon^{\mu\nu}_2=\eta_{\mu\nu}\epsilon^{\mu\nu}_1=\eta_{\mu\nu}\epsilon^{\mu\nu}_2=0
\end{equation}
are imposed. 

This phenomenon is not unique to string theory. In fact, we are very well aware that the same phenomenon occurs in quantum field theory. The S-matrix of one incoming particle state and one outgoing particle state vanishes after imposing the on-shell conditions and the momentum conservation condition, despite the fact that the Green's function of two particles does not vanish. To extract the one-loop renormalization of the kinetic term of fields, one must therefore compute the amputated scattering amplitude by factoring out the polarization tensor and the kinetic term of the field in question, for example, $p^2+m^2$ for a scalar field with mass $m.$ 


To extract the one-loop renormalization of the graviton field, we shall relax the momentum conservation condition and factor out the kinematic factor and the polarization tensor from \eqref{eqn:graviton two point}. For now, for illustrative purposes, let us pretend to relax the on-shell conditions completely. If we expand \eqref{eqn:graviton two point} in $p^2,$ without imposing the on-shell conditions and the momentum conservation condition, we expect to find the form
\begin{equation}
Z=- AV_4g_c^2 K(p_1,p_2,\epsilon_1,\epsilon_2)+\mathcal{O}(p^4)\,,
\end{equation}
where we define
\begin{equation}
K(p_1,p_2,\epsilon_1,\epsilon_2):=\left(p_{1\mu}p_{2\nu}\eta_{\rho\sigma}-p_{1\rho}p_{2\mu}\eta_{\nu\sigma}+\frac{1}{2}p_1\cdot p_2\eta_{\mu\rho}\eta_{\nu\sigma}-\frac{1}{2}p_1\cdot p_2 \eta_{\mu\nu}\eta_{\rho\sigma}\right) \epsilon^{\mu\nu}_1\epsilon^{\rho\sigma}_2\,.
\end{equation}
We shall identify the relation between $A$ and $\delta E$ using the known one-loop correction to the Einstein-Hilbert action from the torus diagram.
This S-matrix can be reproduced by considering the scattering amplitudes between two gravitons whose kinetic term is
\begin{align}\label{eqn:action1}
&\int d^4x \frac{ A}{32\pi^2}\left[ (\partial_\mu h^{\mu\nu})(\partial_\nu h)-(\partial_\mu h^{\rho\sigma}) (\partial_\rho h^{\mu}_{~\sigma})+\frac{1}{2} \eta^{\mu\nu}(\partial_\mu h^{\rho\sigma})(\partial_\nu h_{\rho\sigma})-\frac{1}{2}\eta^{\mu\nu}(\partial_\mu h)(\partial_\nu h)\right]\,.
\end{align}
By using the identification \cite{Polchinski:1998rq}
\begin{equation}
h_{\mu\nu}=-4\pi g_c \epsilon_{\mu\nu} \int \frac{d^4p}{(2\pi)^4}e^{ip\cdot X}\,, 
\end{equation}
we find that the action \eqref{eqn:action1} reproduces the graviton two-point function
\begin{equation}
-Ag_c^2V_4\left(p_{1\mu}p_{2\nu}\eta_{\rho\sigma}-p_{1\rho}p_{2\mu}\eta_{\nu\sigma}+\frac{1}{2}p_1\cdot p_2\eta_{\mu\rho}\eta_{\nu\sigma}-\frac{1}{2}p_1\cdot p_2 \eta_{\mu\nu}\eta_{\rho\sigma}\right) \epsilon^{\mu\nu}_1\epsilon^{\rho\sigma}_2
\end{equation}
As a result, we find the identification 
\begin{equation}\label{eqn:id1}
\delta E=\kappa_4^2\frac{A}{8\pi^2}=\frac{\alpha'}{8\pi}A\,.
\end{equation}

To compute the full kinematic factor $K(p_1,p_2,\epsilon_1,\epsilon_2)$ by evaluating the two-point function $Z,$  we will have to perform a very cumbersome computation. Furthermore, relaxing the on-shell conditions completely may lead to unexpected problems. Therefore, as mentioned before, to compute $Z,$ we shall only relax the momentum conservation condition by setting
\begin{equation}
p_1+p_2=\xi\,,
\end{equation}
and
\begin{equation}
\xi^2=0\,,
\end{equation}
while imposing the on-shell conditions
\begin{equation}
p_1^2=p_2^2=p_1\cdot p_2 =p_{1\mu}\epsilon_1^{\mu\nu}=p_{2\mu}\epsilon^{\mu\nu}_2=\eta_{\mu\nu}\epsilon^{\mu\nu}_1=\eta_{\mu\nu}\epsilon_2^{\mu\nu}=0\,,
\end{equation}
to reduce the full kinematic factor $K(p_1,p_2,\epsilon_1,\epsilon_2)$ to
\begin{equation}
-K(p_1,p_2,\epsilon_1,\epsilon_2):=\eta_{\mu\rho}p_{1\sigma}p_{2\nu} \epsilon_1^{\mu\nu}\epsilon_2^{\rho\sigma}\,.
\end{equation}
With the understanding that we are relaxing the momentum conservation condition, by reading off the form 
\begin{equation}
Z= A V_4 g_c^2 p_{1\rho}p_{2\mu}\eta_{\nu\sigma} \epsilon_1^{\mu\nu}\epsilon_2^{\rho\sigma}\,
\end{equation}
we can therefore compute the one-loop correction to the Einstein-Hilbert action
\begin{equation}
\frac{1}{8\pi^2}\int d^4 x \sqrt{-g} A R\,.
\end{equation}
Note that the same prescription was used in earlier works in the literature \cite{Antoniadis:1996vw,Epple:2004ra,Haack:2015pbv}.

It is extremely important to stress that the hack of relaxing the momentum conservation condition to read off $\delta E$ at this point is a prescription. Therefore, it is necessary to cross-check the results of this draft by different means of computations that do not vanish even with the momentum conservation, i.e. graviton four-point functions. One line of concern one can have is that even in the small warping limit, relaxing the momentum conservation condition can potentially excite the degrees of freedom propagating along the internal directions, which if present can potentially spoil the direct sum structure, due to an unexpected mixing between the external and internal states, that plays an important role in the computations performed in this draft.\footnote{It is important to note that the direct sum structure does not imply that states in the external directions don't interact with the states in the internal directions. What is meant by the direct sum structure is that the states that describe the fluctuations along the external directions are constructed from the free field CFT, and the states that describe the fluctuations along the internal directions are constructed from the Calabi-Yau CFT. In fact, it is very well known that even in Calabi-Yau compactifications at the sphere level scattering amplitudes involving two gravitons and two moduli fields don't vanish despite the fact that the worldsheet CFT for the Calabi-Yau compactification indeed enjoys the direct sum structure! Similarly, it is expected that the string one-loop corrections to the Einstein-Hilbert action will non-trivially depend on moduli, which shall be measured by the moduli dependence of the new supersymmetric index. This implies that if one computes scattering amplitudes involving two gravitons and two moduli for example, one will find a highly non-trivial result that is related to the derivative of the new supersymmetric index with respect to moduli vevs.} The momentum non-conservation is measured by $\xi,$ which we took to be a four-vector that does not have a non-trivial component along the internal direction. Therefore, it is not clear that the prescription we used may cause a deep inelastic scattering where the missing momentum excites the compact degrees of freedom. Furthermore, from the low energy point of view, at the two-derivative level such a coupling between a graviton and an internal degree of freedom, say a modulus, cannot appear. But, na\"{i}ve intuition can fail, and therefore it would be extremely important to confirm this expectation via explicit computations.

To perform a consistency check that the direct product structure of the worldsheet CFT in the weak warping limit is not spoiled when using the momentum relaxation prescription, one can, for example, consider a deep inelastic scattering of an external graviton and an internal graviton, which propagates through the internal dimension.\footnote{We thank the referee for suggesting this computation.} This type of deep inelastic scattering amplitude can be computed for a stack of D-branes in flat spacetime, but it is rather challenging to extend this computation to Calabi-Yau orientifold compactifications because computing scattering amplitudes involving a state propagating through the internal dimension is not very easy. It would be very important to confirm from such a scattering amplitude computation that the internal degrees of freedom are not excited in the graviton scattering amplitudes, and we are actively investigating this issue. We hope to come back to this issue with more concrete computation in the near future.  

As the computations performed in this draft are new, there is no known direct cross-check of the prescription. But, in simpler settings, one can perform some cross-checks. In type I compactifications on $K3\times T^2,$ the same prescription was used in \cite{Antoniadis:1996vw} to compute the graviton two-point function at the string one-loop. The results of \cite{Antoniadis:1996vw} were reproduced by the graviton three-point function in section 3 of  \cite{Antoniadis:2002tr}. As an extra cross-check of the prescription in a simpler setting, in \S\ref{sec:torus} we will compute the graviton two-point function on the worldsheet torus and obtain the result that is in perfect agreement with the well-known result in the literature \cite{Kiritsis:1994ta,Antoniadis:1996vw,Antoniadis:1997eg,Antoniadis:2002tr,Green:1997di,Green:1997as,Kiritsis:1997em,Russo:1997mk,Becker:2002nn,Liu:2022bfg}.

Finally, we shall now clarify what we mean by computing the string one-loop corrections to the EH action in this draft. Even though the graviton correlation function can be evaluated using the rules of the free field CFT, within the regime of the validity of our approximation scheme, the graviton two-point functions still depend on the spectrum of the internal CFT which is beyond reach in general. And to evaluate the moduli dependence of the string one-loop corrections to the EH action, it is crucial to have some access to the spectrum of the internal CFT.

So, na\"{i}vely, one might conclude that we reached a hard wall even after invoking several approximations and the prescription to make progress because our answer depends on the spectrum of the internal CFT. Furthermore, because the Einstein-Hilbert action cannot be written as an F-term in the effective supergravity, it seems hard to imagine a possibility that the EH action is related to a computable protected quantity.

Despite this expectation, quite surprisingly, in the subsequent sections we will show that the string one-loop corrections to the Einstein-Hilbert action are determined in terms of the new supersymmetric index and the Witten index of various one-loop diagrams. Therefore, despite the na\"{i}ve expectation, the one-loop diagrams only depend on more computable crude details of the string spectrum. In \S\ref{sec:discussion}, we shall explain why this structure is at some level expected by comparing the graviton two point function to topological string theory amplitudes at one-loop.

In the subsequent sections, we will evaluate contributions to $\delta E$ from Riemann surfaces of genus one with and without boundaries.
\subsection{Torus diagram}\label{sec:torus}
In this section, we shall review the well-known corrections to the Einstein-Hilbert action from the torus diagram. Relevant papers include \cite{Kiritsis:1994ta,Antoniadis:1996vw,Antoniadis:1997eg,Antoniadis:2002tr,Green:1997di,Green:1997as,Kiritsis:1997em,Russo:1997mk,Becker:2002nn,Liu:2022bfg}. For the torus diagram, there are two spin structures for left and right-moving fermions. We need to sum over those two spin structures accordingly. Because the Einstein-Hilbert term is CP even, one needs to sum over (even,even) and (odd,odd) spin structures. 

As in the heterotic compactifications, the sum over (even,even) spin structures yields zero \cite{Kiritsis:1994ta,Antoniadis:1996vw}. This can be also understood as a corollary of \eqref{eqn:vanishing int}. The only non-vanishing contribution, therefore, comes from the (odd,odd) spin structure which yields \cite{Antoniadis:1997eg}
\begin{equation}\label{eqn:torus contribution}
\delta E_\mathcal{T}=\pm \frac{1}{2^5\cdot 3\cdot \pi} \text{Tr}_{R,R}\left((-1)^{F_L+F_R}q^{L_0-\frac{3}{8}}\bar{q}^{\bar{L}_0-\frac{3}{8}}\right)_{int}^{closed}\,,
\end{equation}
where $+$ corresponds to type IIB and $-$ correponds to type IIA. This result was obtained by evaluating the graviton three-point function on the worldsheet torus \cite{Antoniadis:1997eg}. Note that
\begin{equation}
\text{Tr}_{R,R}\left((-1)^{F_L+F_R}q^{L_0-\frac{3}{8}}\bar{q}^{\bar{L}_0-\frac{3}{8}}\right)_{int}^{closed}
\end{equation}
is the Witten index \cite{Witten:1982df} of the compactification manifold. In a geometric phase, we have
\begin{equation}
\delta E_\mathcal{T}=\pm \frac{1}{2^5\cdot 3\cdot \pi}\chi\,,
\end{equation}
where $\chi$ is the Euler characteristic of the Calabi-Yau threefold. 

This result can be compared to the $\mathcal{O}(\alpha'^3 R^4)$ terms obtained in 10d type II string theory \cite{Green:1997di,Green:1997as,Kiritsis:1997em,Russo:1997mk}. Dimensional reduction of the $\mathcal{O}(\alpha'^3 R^4)$ terms yield the corrections to the Einstein-Hilbert action in the 4d effective supergravity. This result can be extracted by $\zeta(3)$ with $g_s^2\pi^2/3$, for example, in eqn. (A.21) in \cite{Becker:2002nn}, or by extracting the string one-loop correction from the eqn (D.20) in \cite{Liu:2022bfg}. For convenience let us take type IIB string theory. We write (D.20) in \cite{Liu:2022bfg} here
\begin{equation}
-\frac{1}{2^7 \pi^7 \alpha'^4}\int d^4 x \left[ 1536 \alpha (2\pi)^3 \chi f_0\right]R\,,
\end{equation}
where we define $\alpha:= (\alpha')^3/(3\cdot 2^{12}),$ and $f_0$ the non-holomorphic Eisenstein series of weight $3/2$ whose expansion is given as $f_0=2\zeta(3) e^{-3\phi/2}+\frac{2\pi^2}{3} e^{\phi/2}+\dots.$ By expanding the numerical factors, we get
\begin{equation}
-\frac{1}{2^6\pi^4\alpha'} \int d^4 x (\zeta(3)e^{-3\phi/2}+e^{\phi/2}\pi^2/3+\dots) \chi R\,.
\end{equation}
As a result, we can extract the torus contribution
\begin{equation}
\delta E_{\mathcal{T}}=-\frac{\chi\kappa_4^2}{3\cdot 2^5\cdot \pi^2\alpha'}=-\frac{\chi}{3\cdot 2^5\cdot \pi}\,,
\end{equation}
which agrees with \eqref{eqn:torus contribution}.

In order to crosscheck the momentum non-conservation prescription, we will directly compute the (odd,odd) spin structure amplitude. In the (odd,odd) spin structure, due to the presence of the non-trivial superconformal Killing vectors, we must fix $\theta$ and $\bar{\theta}$ of some vertex operators to zero. This forces some of the vertex operators to have non-trivial picture numbers. Furthermore, we shall include the PCOs. There are in total four different ways to distribute the picture numbers
\begin{equation}\label{eqn:picture 1}
\langle V_g^{(-1,-1)}(z_1,p_1,\epsilon_1)V_g^{(0,0)}(z_2,p_2,\epsilon_2)\rangle \,,\quad \langle V_g^{(0,0)}(z_1,p_1,\epsilon_1)V_g^{(-1,-1)}(z_2,p_2,\epsilon_2)\rangle\,,
\end{equation}
\begin{equation}\label{eqn:picture 2}
\langle V_g^{(-1,0)}(z_1,p_1,\epsilon_1)V_g^{(0,-1)}(z_2,p_2,\epsilon_2)\rangle\,,\quad\langle V_g^{(0,-1)}(z_1,p_1,\epsilon_1)V_g^{(-1,0)}(z_2,p_2,\epsilon_2)\rangle\,.
\end{equation}
Because gravitons are massless \eqref{eqn:picture 1} vanish.

We shall distribute the picture numbers as the first in \eqref{eqn:picture 2} 
\begin{equation}
K_\sigma^{(o)}(p_1,p_2):= \langle V_g^{(-1,0)}(z_1,p_1)V_g^{(0,-1)} (z_2,p_2) e^{\phi}T_F (z_0)  e^{\bar{\phi}} \bar{T}_F (\bar{z}_0)\rangle\,,
\end{equation}
where $ e^{\phi} T_F(z_0)$ and $e^{\bar{\phi}}\bar{T}_F(\bar{z}_0)$ are due to the insertion of PCOs which we normalize as
\begin{equation}
\mathcal{X}=\{ Q_B,\xi(z)\}= e^\phi T_F+\dots\,,
\end{equation}
and we normalize $T_F= i(2/\alpha')^{1/2}\psi \cdot \partial X+\dots.$\footnote{Note that we used the standard normalization of ghost fields following \cite{Polchinski:1998rq}, which resulted in a difference of the overall normalization by 2 from \cite{Sen:2021tpp,Alexandrov:2021shf}.} Because we are in the (odd,odd) spin structure, there are four fermionic zero modes both from left and right moving sectors of the non-compact part of the worldsheet CFT. Therefore, we should soak up the fermionic zero modes to have a non-vanishing result by pairing the zero-mode integral with fermion insertions. We shall carefully demonstrate this procedure as follows. The component form of the correlator is
\begin{align}
\mathfrak{K}^{(o)}:=&- \frac{2^2g_c^2}{\alpha'^2} (2\pi)^2\epsilon_{\mu\nu}^1\epsilon_{\rho\sigma}^2\left\langle e^{-\phi}\psi_1^\mu \left(i\bar{\partial}X_1^\nu +\frac{\alpha'}{2}p_1\cdot \bar{\psi}_1\bar{\psi}_1^\nu\right) \left(i\partial X_2^\rho +\frac{\alpha'}{2}p_2\cdot \psi_2\psi_2^\rho\right)e^{-\bar{\phi}}\bar{\psi}^\sigma \right.\nonumber\\
&\qquad\qquad\qquad\times\left(e^{\phi}\psi\cdot\partial X\right)(z_0) \left(e^{\bar{\phi}}\bar{\psi}\cdot\bar{\partial} X\right)(\bar{z}_0) \biggr\rangle \,.
\end{align}
First, we compute
\begin{equation}
\langle \psi_1^\mu p_2\cdot \psi_2\psi_2^\rho \psi^\delta\rangle= p_{2\alpha}\eta(\tau)^4 \epsilon^{\mu\alpha\rho\delta}\,.
\end{equation}
Similarly, we compute
\begin{equation}
\langle p_1 \cdot \bar{\psi}_1\bar{\psi}_1^\nu \bar{\psi}^\sigma \bar{\psi}^{\epsilon}\rangle =p_{1\beta} (\eta(\tau)^4)^*\epsilon^{\beta\nu\sigma\epsilon}\,.
\end{equation}
From the $b,c$ ghost system we obtain $|\eta(\tau)|^4,$ and we have an additional factor $(2\pi)^2$ in the vacuum amplitude\footnote{Note that this factor was introduced to take into account that the vertex position convention used in this paper is different from the vertex position convention of \cite{Polchinski:1998rq}. In this note, we normalized $z$ such that $z\sim z+1\sim z+\tau.$ If we used the convention of \cite{Polchinski:1998rq}, we wouldn't have the additional $(2\pi)^2$ in the vacuum amplitude. Instead, the vertex position integral will give $((2\pi)^2 \tau_2)^2,$ instead of $\tau_2^2,$ and the contraction between bosons in the PCOs will generate $-\alpha'/(8\pi \tau_2),$ instead of $-\pi\alpha'/2\tau_2,$ which in total yield the same answer.} and from the $\beta,\gamma$ ghost system we obtain $|\eta(\tau)|^{-4}.$ The boson partition function generates $|\eta(\tau)|^{-8}$ factor. Integral over the closed string momentum along the non-compact directions generates
\begin{equation}
\int \frac{d^4k}{(2\pi)^4}e^{-\pi \tau_2\alpha' k^2} =\frac{1}{2^4\pi^4\alpha'^2\tau_2^2}\,.
\end{equation}
The contraction between the bosonic components in the PCOs generate $-\frac{\pi\alpha'}{2\tau_2}.$ The internal part of the CFT produces 
\begin{equation}
\chi:=\text{Tr}_{R,R}^{(int)}\left((-1)^{F+\bar{F}}q^{L_0-\frac{3}{8}}\bar{q}^{\bar{L}_0-\frac{3}{8}}\right)\,.
\end{equation}
Integral over the vertex moduli generates $4\tau_2^2.$ Finally, we need to include the volume factor $V_4.$ Combining all these, we obtain
\begin{equation}
\mathfrak{K}^{(o)}=\pm \frac{V_4g_c^2}{2\pi\alpha'}\int \frac{d^2\tau}{2\tau_2^2}\epsilon_{\mu\nu}^1\epsilon_{\rho\sigma}^2 p_{1\alpha}p_{2\beta} \epsilon^{\mu\alpha\rho}\epsilon^{\beta\nu\sigma} \chi\,,
\end{equation}
where the overall sign $+$ is for type IIA and $-$ is for type IIB. Using the identity
\begin{equation}
\int \frac{d^2\tau}{\tau_2^2}=\frac{\pi}{3}\,,
\end{equation}
we obtain
\begin{equation}
\mathfrak{K}^{(o)}= \pm\frac{1}{3}\frac{V_4g_c^2}{2^2\alpha'}\epsilon_{\mu\nu}^1\epsilon_{\rho\sigma}^2 p_{1\alpha}p_{2\beta} \epsilon^{\mu\alpha\rho}\epsilon^{\beta\nu\sigma} \chi\,.
\end{equation}
After rewriting the Levi-Civita symbols, we obtain
\begin{equation}\label{eqn:torus res}
\mathfrak{K}^{(o)}=\mp\frac{1}{3}\frac{V_4g_c^2}{2^2\alpha'}\epsilon_{\mu\nu}^1\epsilon_{\rho\sigma}^2 p_1^\sigma p_2^\mu \eta^{\nu\rho}\chi\,.
\end{equation}
Here we now obtain 
\begin{equation}
A=\pm\frac{1}{3} \frac{1}{2^2\alpha'} \chi\,,
\end{equation}
which corresponds to
\begin{equation}
\delta E_\mathcal{T}=\pm \frac{1}{2^5\cdot 3\cdot \pi } \chi\,.
\end{equation}
As a result, we find the perfect agreement with the known result.
\subsection{Annulus}
In this section, we compute the contribution from annuli diagrams $Z_A$ to $Z.$ We have
\begin{align}\label{eqn:annulus 1}
Z_A=&\frac{V_4}{2^8 \pi^4 \alpha'^2} \sum_{\alpha,\beta \text{ even}}\int_0^\infty \frac{dt}{t^3} \int_A d^2 z_1 \int_A  d^2z_2 \biggl[ \langle V_g^{(0,0)}(z_1,p_1,\epsilon_1) V_g^{(0,0)}(z_2,p_2,\epsilon_2) \rangle_A^s   \nonumber\\
&\qquad\qquad\qquad\qquad\times (-1)^{\alpha+\beta}\frac{\vartheta_{\alpha,\beta}(\tau)}{\eta(\tau)^3}\text{Tr}_{\alpha}\left((-1)^{\beta F} q^{L_0-\frac{3}{8}} \right)^{open}_{int} \biggr]\,,
\end{align}
where the factor $1/(2^6\pi^4t^2)$ was induced due to the integral over the momentum of the open string
\begin{equation}
\int \frac{d^4 k}{(2\pi)^4} e^{-2\pi t\alpha' k^2}=\frac{1}{2^6\pi^4\alpha'^2 t^2}\,,
\end{equation}
and the factor $(-1)^{\alpha+\beta}\vartheta_{\alpha,\beta}(\tau)/\eta(\tau)^3$ comes from the non-compact directions and the ghost system, and we included an additional factor $1/2$ due to the GSO projection. Unlike in the case of toroidal amplitude, we don't include an additional factor of $(2\pi)^2$ which was introduced to take into account the fact that the ambient torus has the periodicity $z\sim z+1\sim z+\tau.$ The reason why we don't have to do so here is because the fermion correlator contains the explicit factor of $(2\pi)^2,$ to automatically take this convention into account.\footnote{Said differently, in the toroidal amplitude computation, we had two integral over the vertex operator positions, and the two-point correlators of the vertex operators only produced $1/\text{vol}$ factor, instead of $1/\text{vol}^2.$ This is because the fermion terms in the vertex operators were used to soak up fermion zero modes. When changing the convention from that of \cite{Polchinski:1998rq} to our convention, one has to correctly change the normalization of the fermion zero modes. } Note that we included the multiplicities due to the Chan-Paton factors in the trace for the internal CFT. The sum runs over even spin structures because the Einstein-Hilbert term is CP-even. Using \eqref{eqn:graviton correlators}, we find
\begin{align}
\langle V_g^{(0,0)}(z_1,p_1,\epsilon_1)V_g^{(0,0)}(z_2,p_2,\epsilon_2)\rangle_A^{s}=&-g_c^2 \eta_{\mu\rho}p_{1\sigma}p_{2\nu}\epsilon_1^{\mu\nu}\epsilon_2^{\rho\sigma}\bigg[\langle \partial X_1\partial X_2\rangle_A (\langle \bar{\psi}_2\bar{\psi}_1\rangle_A^s)^2\nonumber\\
& +\langle \partial X_1\bar{\partial}X_2\rangle_A(\langle \psi_2\bar{\psi}_1\rangle_A^s)^2+c.c\bigg]\,,
\end{align}
where we used the shorthand notation for the spin structure $s=(\alpha,\beta).$ For example, a trace in the spin structure $s$ is written as
\begin{equation}
\text{Tr}_{\alpha}\left((-1)^{\beta F}q^{L_0-\frac{3}{8}}\right)\,,
\end{equation}
where $\alpha=0$ stands for the NS sector, and $\alpha=1$ stands for the R sector. It is important to note that the bosonic two-point functions do not depend on the spin structure, but the fermionic two-point functions do as we have for the spin structure $s=(\alpha,\beta)$ \eqref{eqn:fermion on sigma}
\begin{equation}\label{eqn: ferm identity imp0}
(\langle\bar{\psi}_2(0)\bar{\psi}_1(z)\rangle_A^s)^2=\left(\frac{\vartheta_{\alpha,\beta}(z|\tau)\vartheta_{1,1}'(0|\tau)}{\vartheta_{\alpha,\beta}(0|\tau)\vartheta_{1,1}(z|\tau)}\right)^2\,.
\end{equation}
To evaluate the integral over the vertex positions, we will use the identities \cite{Alexandrov:2022mmy}
\begin{equation}\label{eqn:identity imp0}
\sum_{\alpha,\beta \text{ even}} (-1)^{\alpha+\beta}\vartheta_{\alpha,\beta}(0|\tau)\text{Tr}_{\alpha}\left((-1)^{\beta F} q^{L_0-\frac{3}{8}} \right)_{int}=0\,,
\end{equation}
and \eqref{eqn:identity imp1}
\begin{equation}\label{eqn:identity imp2}
\left(\frac{\vartheta_{\alpha,\beta}(z|\tau)\vartheta_{1,1}'(0|\tau)}{\vartheta_{\alpha,\beta}(0|\tau)\vartheta_{1,1}(z|\tau)}\right)^2=\frac{\vartheta_{\alpha,\beta}''(0|\tau)}{\vartheta_{\alpha,\beta}(0|\tau)}-\partial_z^2\log\vartheta_{1,1}(z|\tau)\,.
\end{equation}
Because the term $\partial_z^2\log\vartheta_{1,1}(z|\tau)$ does not depend on the spin structure, after summing over the spin structures $(\alpha,\beta)$ the contribution from $\partial_z^2\log\vartheta_{1,1}(z|\tau)$ cancels out. Therefore, we can rewrite the integral as
\begin{align}
\delta E_A:=&\frac{\alpha'  Z_A}{2^3\pi V_4 g_c^2 \eta_{\mu\rho}p_{1\sigma}p_{2\nu}\epsilon_1^{\mu\nu}\epsilon_2^{\rho\sigma}}\,,\\
=&-\frac{1}{2^{11}\pi^5\alpha'}\sum_{\alpha,\beta\text{ even}} \int_0^\infty\frac{dt}{t^3}\int_A d^2 z_1 \int_A  d^2z_2\biggl[\left(\langle\partial X_1\partial X_2\rangle_A-\langle \partial X_1\bar{\partial}X_2\rangle_A+c.c\right) \nonumber\\
&\qquad\qquad\qquad\qquad\qquad\qquad\times(-1)^{\alpha+\beta}\frac{\vartheta_{\alpha,\beta}''(0|\tau)}{\eta(\tau)^3}\text{Tr}_{\alpha}\left((-1)^{\beta F} q^{L_0-\frac{3}{8}} \right)_{int}^{open} \biggr]\,,\label{eqn:delta E intermediate0}\\
=&-\frac{1}{2^{12}\pi^4}\sum_{\alpha,\beta \text{ even}} \int_0^\infty \frac{dt}{t^2} (-1)^{\alpha+\beta} \frac{\vartheta_{\alpha,\beta}''(0|\tau)}{\eta(\tau)^3}\text{Tr}_{\alpha}\left((-1)^{\beta F} q^{L_0-\frac{3}{8}} \right)^{open}_{int} \,,\label{eqn:delta E intermediate}
\end{align}
where we have used the identity \eqref{eqn:boson on sigma} to perform the integral over the vertex positions.

Now that we reduced the integral, we will use the identities \eqref{eqn:identity imp0} and \eqref{eqn:identity imp2} again to rewrite $\vartheta_{\alpha,\beta}''(0|\tau)$ in a more useful form. In principle, one can leave $z$ in \eqref{eqn:identity imp2} arbitrary as long as $\vartheta_{1,1}(z|\tau)\neq0.$ But, we will fix $z=\frac{\tau}{2}$ to rewrite $\vartheta_{\alpha,\beta}''(0|\tau)$ as
\begin{equation}
\vartheta_{\alpha,\beta}''(0|\tau)=\vartheta_{\alpha,\beta}(0|\tau)\partial_z^2\log\vartheta_{1,1}(z|\tau)|_{z=\tau/2}+4\pi^2  \frac{\vartheta_{\alpha,\beta}(\frac{\tau}{2}|\tau)^2\eta(\tau)^6}{\vartheta_{\alpha,\beta}(0|\tau)\vartheta_{1,1}(\frac{\tau}{2}|\tau)^2}\,,
\end{equation}
and rewrite $\delta E_A$ as
\begin{equation}
\delta E_A=-\frac{1}{2^9\pi^2}\sum_{\alpha,\beta\text{ even}} \int_0^\infty \frac{dt}{2t^2} (-1)^{\alpha+\beta}\frac{\vartheta_{\alpha,\beta}(\frac{\tau}{2}|\tau)^2\eta(\tau)^3}{\vartheta_{\alpha,\beta}(0|\tau)\vartheta_{1,1}(\frac{\tau}{2}|\tau)^2}\text{Tr}_{\alpha}\left((-1)^{\beta F} q^{L_0-\frac{3}{8}} \right)^{open}_{int} \,.
\end{equation}
Because of the term $\vartheta_{\alpha,\beta}(\frac{\tau}{2}|\tau)$ in the numerator, contribution from the spin structure $(\alpha,\beta)=(0,1)$ vanishes. Therefore, we only need to study the spin structures 
\begin{equation}
(\alpha,\beta)=(0,0)\,,~(1,0)\,.
\end{equation} 

Let us study a contribution from a massive state with $(h,Q).$ Using the identity
\begin{equation}
\frac{\vartheta_{\alpha,0}(\frac{\tau}{2}|\tau)^2}{\vartheta_{1,1}(\frac{\tau}{2}|\tau)^2}= -\frac{\vartheta_{\alpha+1,0}(\tau)^2}{\vartheta_{0,1}(\tau)^2}\,,
\end{equation}
we reorganize the integrand as
\begin{align}
\mathfrak{Z}_A^{(h,Q)}:=&\sum_{\alpha,\beta\text{ even}} (-1)^{\alpha+\beta}\frac{\vartheta_{\alpha+1,\beta}(0|\tau)^2\eta(\tau)^3}{\vartheta_{\alpha,\beta}(0|\tau)\vartheta_{0,1}(0|\tau)^2}\text{Tr}_{\alpha}^{(h,Q)}\left((-1)^{\beta F} q^{L_0-\frac{3}{8}} \right)^{open}_{int} \,,\\
=&\frac{1}{\vartheta_{0,1}(\tau)^2} q^{h-\frac{1+Q}{4}} \left(\vartheta_{Q,0}(2\tau)\vartheta_{1,0}(\tau)^2-\vartheta_{1-Q,0}(2\tau)\vartheta_{0,0}(\tau)^2\right)\,.
\end{align}
Let us use the addition rules \eqref{eqn:addition rule1}, \eqref{eqn:addition rule3}, and \eqref{eqn:addition rule4} to rewrite $\mathfrak{Z}_A^{(h,Q)}$ as 
\begin{align}
\mathfrak{Z}_A^{(h,Q)}=&q^{h-\frac{1+Q}{4}}\frac{2\vartheta_{Q,0}(2\tau)\vartheta_{0,0}(2\tau)\vartheta_{1,0}(2\tau)-\vartheta_{1-Q,0}(2\tau)(\vartheta_{0,0}(2\tau)^2+\vartheta_{1,0}(2\tau)^2)}{\vartheta_{0,0}(2\tau)^2-\vartheta_{1,0}(2\tau)^2}\,,\\
=&(-1)^{Q}q^{h-\frac{1+Q}{4}}\vartheta_{1-Q,0}(2\tau)\,.
\end{align}
Finally, by using the identity \eqref{eqn:the new supersymmetric index id}, we conclude
\begin{equation}\label{eqn:new index annulus 1}
\mathfrak{Z}_A^{(h,Q)}= \text{Tr}_R^{(h,Q)}\left((-1)^{F-\frac{3}{2}}F q^{L_0-\frac{3}{8}}\right)^{open}_{int}\,.
\end{equation}

Let us now study contributions from the massless states. We shall start with the vacuum state. As the character for the vacuum state can be written as
\begin{equation}
\text{Tr}_{\alpha}^{(vac)}\left((-1)^{\beta F} q^{L_0-\frac{3}{8}} \right)^{open}_{int} =\text{Tr}_{\alpha}^{(0,0)}\left((-1)^{\beta F} q^{L_0-\frac{3}{8}} \right)^{open}_{int}-\text{Tr}_{\alpha}^{(\frac{1}{2},1)}\left((-1)^{\beta F} q^{L_0-\frac{3}{8}} \right)^{open}_{int}\,,
\end{equation}
we conclude that the vacuum contribution is
\begin{align}
\mathfrak{Z}_A^{(vac)}:=&\sum_{\alpha,\beta\text{ even}} (-1)^{\alpha+\beta}\frac{\vartheta_{\alpha+1,\beta}(0|\tau)^2\eta(\tau)^3}{\vartheta_{\alpha,\beta}(0|\tau)\vartheta_{0,1}(0|\tau)^2}\text{Tr}_{\alpha}^{(vac)}\left((-1)^{\beta F} q^{L_0-\frac{3}{8}} \right)^{open}_{int} \,,\\
=&\text{Tr}_R^{(vac)}\left((-1)^{F-\frac{3}{2}}F q^{L_0-\frac{3}{8}}\right)^{open}_{int}\,. \label{eqn:new index annulus 2}
\end{align}
Let us finally study the $(\pm)$ states. By using the identity \eqref{eqn:massless characters2}, we find
\begin{equation}
\text{Tr}_{\alpha}^{(\pm)} \left((-1)^{\beta F} q^{L_0-\frac{3}{8}} \right)^{open}_{int}=\frac{1}{2} \text{Tr}_{\alpha}^{(\frac{1}{2},1)}\left((-1)^{\beta F} q^{L_0-\frac{3}{8}} \right)^{open}_{int}\,,
\end{equation}
for the even spin structures $(s_1,s_2),$ and
\begin{equation}
\text{Tr}_R^{(\pm)}\left((-1)^{F-\frac{3}{2}}F q^{L_0-\frac{3}{8}}\right)^{open}_{int}=\frac{1}{2}\text{Tr}_R^{(\frac{1}{2},1)}\left((-1)^{F-\frac{3}{2}}F q^{L_0-\frac{3}{8}}\right)^{open}_{int}\mp \frac{3}{2}\,.
\end{equation}
As a result, we conclude
\begin{align}
\mathfrak{Z}_A^{(\pm)}:=&\sum_{\alpha,\beta\text{ even}} (-1)^{\alpha+\beta}\frac{\vartheta_{\alpha+1,\beta}(0|\tau)^2\eta(\tau)^3}{\vartheta_{\alpha,\beta}(0|\tau)\vartheta_{0,1}(0|\tau)^2}\text{Tr}_{\alpha}^{(\pm)}\left((-1)^{\beta F} q^{L_0-\frac{3}{8}} \right)^{open}_{int} \,,\\
=&\text{Tr}_R^{(\pm)}\left((-1)^{F-\frac{3}{2}}F q^{L_0-\frac{3}{8}}\right)^{open}_{int}\pm \frac{3}{2}\,. \label{eqn:new index annulus 3}
\end{align}
Note that we used \eqref{eqn:massless f}.

Combining \eqref{eqn:new index annulus 1}, \eqref{eqn:new index annulus 2}, and \eqref{eqn:new index annulus 3}, we arrive at one of the main results of this paper
\begin{equation}\label{eqn:main 1}
\boxed{\delta E_A=\frac{1}{2^9\pi^2}\int_0^\infty \frac{dt}{2t^2} \Bigl[\text{Tr}_R \left((-1)^{F-\frac{3}{2}}F q^{L_0-\frac{3}{8}}\right)^{open}_{int}+\frac{3}{2} \left(n^+_A-n^-_A\right)\Bigr]\,,}
\end{equation}
where $n^{(\pm)}_A$ is a number of the $(\pm)$ state. Note that one can use the following identity 
\begin{equation}
\text{Tr}_R\left((-1)^{F-\frac{3}{2}} q^{L_0-\frac{3}{8}}\right)_{int}^{open}=n_A^+-n_A^-\,,
\end{equation}
to rewrite \eqref{eqn:main 1} as
\begin{equation}
\delta E_A=\frac{1}{2^{9}\pi^2}\int_0^\infty \frac{dt}{2t^2} \text{Tr}_R \Bigl[(-1)^{F-\frac{3}{2}}\left(F+\frac{3}{2}\right) q^{L_0-\frac{3}{8}}\Bigr]^{open}_{int}\,.
\end{equation}
\subsection{M{\"o}bius strip and Klein bottle}
In this section, we study the contributions from the M{\"o}bius strip and the Klein bottle. Most of the computation goes through the same as before.

We first study the M{\"o}bius strip contribution. We note that for the M{\"o}bius strip $\tau_\mathcal{M}$ is now
\begin{equation}
\tau_\mathcal{M}=\frac{1}{2}+it\,,
\end{equation}
and the orientifold projection $\Omega$ is inserted in the trace. As a result, we obtain
\begin{equation}\label{eqn:main 2}
\boxed{\delta E_\mathcal{M}=\frac{1}{2^9\pi^2}\int_0^\infty \frac{dt}{2t^2} \Bigl[\text{Tr}_R \left((-1)^{F-\frac{3}{2}}F\Omega q^{L_0-\frac{3}{8}}\right)^{open}_{int}+\frac{3}{2} \left(n^+_\mathcal{M}-n^-_\mathcal{M}\right)\Bigr]\,.}
\end{equation}

To read off the contribution from the Klein bottle, we should first remark on a few important details. The Klein bottle is obtained by orientifolding closed string, torus. We set the torus modulus to be
\begin{equation}
\tau_\mathcal{K}=2it\,,
\end{equation}
and
\begin{equation}
I_\mathcal{K}(z)=1-\bar{z}+\frac{\tau_\mathcal{K}}{2}\,.
\end{equation}
To obtain the correction to the Einstein-Hilbert action, we must sum over (even,even) and (odd,odd) spin structures. Because the left-handed spin structure is identical to the right-handed spin structure, we can focus only on the left-handed spin structure. The contribution from the (even,even) spin structure can be read off from the annulus contribution by replacing $\tau_A$ with $\tau_\mathcal{K}$ and inserting the orientifold projection $\Omega.$ Note that the momentum integral now yields $1/ (2^4\pi^2\alpha'^2t^2)$ instead of $1/(2^6\pi^2\alpha'^2t^2)$ \cite{Polchinski:1998rq}. Additionally, we shall include a factor of $2$ because $\im\tau_\mathcal{K}$ is twice that of the annulus. 
\begin{equation}\label{eqn:main 3}
\boxed{\delta E_\mathcal{K}^{(e,e)}=\frac{1}{2^6\pi^2}\int_0^\infty \frac{dt}{2t^2}\Bigl[ \text{Tr}_{R,R} \left((-1)^{F}F\Omega q^{L_0-\frac{3}{8}}\bar{q}^{\bar{L}_0-\frac{3}{8}}\right)^{closed}_{int}+\frac{3}{2} \left(n^+_\mathcal{K}-n^-_\mathcal{K}\right)\Bigr]\,.}
\end{equation}
Similarly, the contribution from the (odd,odd) spin structure can be read off from \eqref{eqn:torus contribution} by inserting the orientifold projection $\Omega$
\begin{equation}\label{eqn:main 4}
\boxed{\delta E_{\mathcal{K}}^{(o,o)}=\frac{1}{2^5\cdot 3\cdot \pi} \text{Tr}_{R,R}\left((-1)^{F}\Omega q^{L_0-\frac{3}{8}}\bar{q}^{\bar{L}_0-\frac{3}{8}}\right)^{closed}_{int}\,.}
\end{equation}
In a geometric phase, we have \cite{Brunner:2003zm}
\begin{equation}
\delta E_{\mathcal{K}}^{(o,o)}=\frac{1}{2^5\cdot 3\cdot \pi} \chi_f\,,
\end{equation}
where we define
\begin{equation}
\chi_f:= \sum_{p,q} (-1)^{p+q} (h_+^{p,q}-h_-^{p,q})\,,
\end{equation}
and $h_{\pm}^{p,q}$ are orientifold even (odd) hodge numbers.

By combining \eqref{eqn:main 1}, \eqref{eqn:main 2}, \eqref{eqn:main 3}, and \eqref{eqn:main 4}, we arrive at the one-loop correction to the EH term
\begin{equation}\label{eqn:main 5}
\boxed{\delta E= \frac{1}{2} \left(\delta E_\mathcal{T}+\delta E_A+\delta E_\mathcal{M}+\delta E_\mathcal{K}\right)\,.}
\end{equation} 

\section{The one-loop correction and small cycles}\label{sec:small cycles}
In this section, we shall study the size of the one-loop correction $\delta E$ and its effect on the one-loop correction to the K{\"a}hler potential. 

First, we shall study the contribution from a massive state with $(h,Q).$ Let us define
\begin{equation}
\mathfrak{E}^{(h,Q)}_\sigma:= c_\sigma(-1)^Q\int_1^\infty \frac{dt}{2t^2}\text{Tr}_R^{(h,Q)}\left((-1)^{F-\frac{3}{2}}F q^{L_0-\frac{3}{8}}\right)\,,
\end{equation}
where $c_{T,\mathcal{M}}=2^{-9}\pi^{-2}$ and $c_{\mathcal{K}}=2^{-6}\pi^{-2}.$ A few comments are in order. $\mathfrak{E}_\sigma^{(h,Q)}$ can be understood as a contribution to $\delta E_\sigma$ from an irreducible state with $(h,Q).$  To compute $\delta E_\sigma,$ one should integrate $t$ between $0$ and $\infty$ as
\begin{equation}
\delta E_\sigma=-c_\sigma\int_0^\infty \frac{dt}{2t^2} \text{Tr}_{R,R} \left((-1)^{F}F\Omega q^{L_0-\frac{3}{8}}\bar{q}^{\bar{L}_0-\frac{3}{8}}\right)\,.
\end{equation}
But, to define $\mathfrak{E}^{(h,Q)},$ we only integrated $t$ from $1$ to $\infty.$ The reason for doing so is to isolate the open string channel and the most sizable contribution therein, the IR contribution. The contributions from the UV region when $t<1$ won't be very important for our discussion as the most important corrections are coming from states with low energy. Note further that the UV divergence is canceled by requiring the tadpole cancellation.

\begin{figure}
\center
\includegraphics[width=7cm]{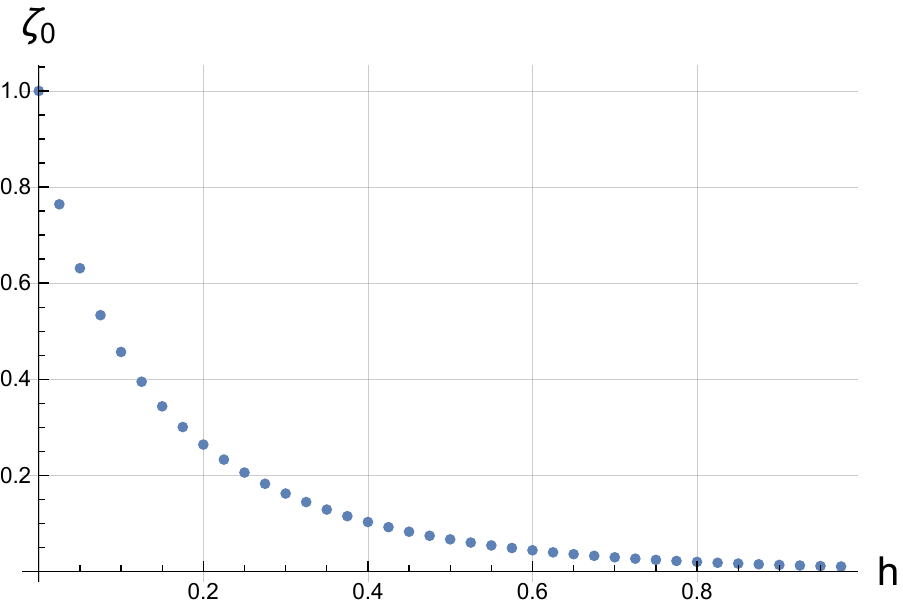}
\includegraphics[width=7cm]{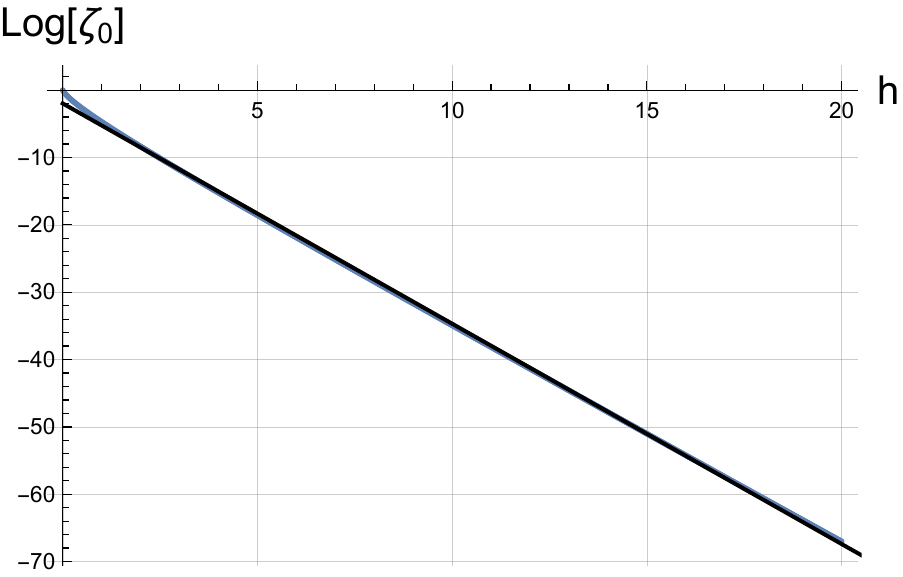}
\caption{$\zeta_0:= \mathfrak{E}_\sigma^{(h,0)}/c_\sigma$ for massive states with various values of $h.$ Left: We plot $\zeta_0$ for small $h.$ Right: Blue dots represent the numerical value of $\log(\zeta_0)$ at given $h.$ The black line is a linear fit for $\log(\zeta_0).$}\label{figure:1}
\end{figure}

\begin{figure}
\center
\includegraphics[width=7cm]{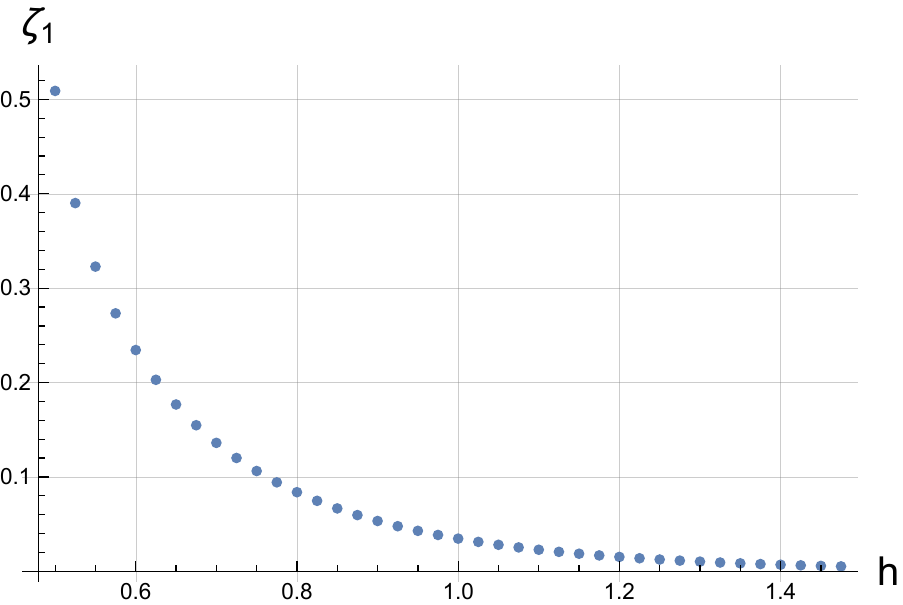}
\includegraphics[width=7cm]{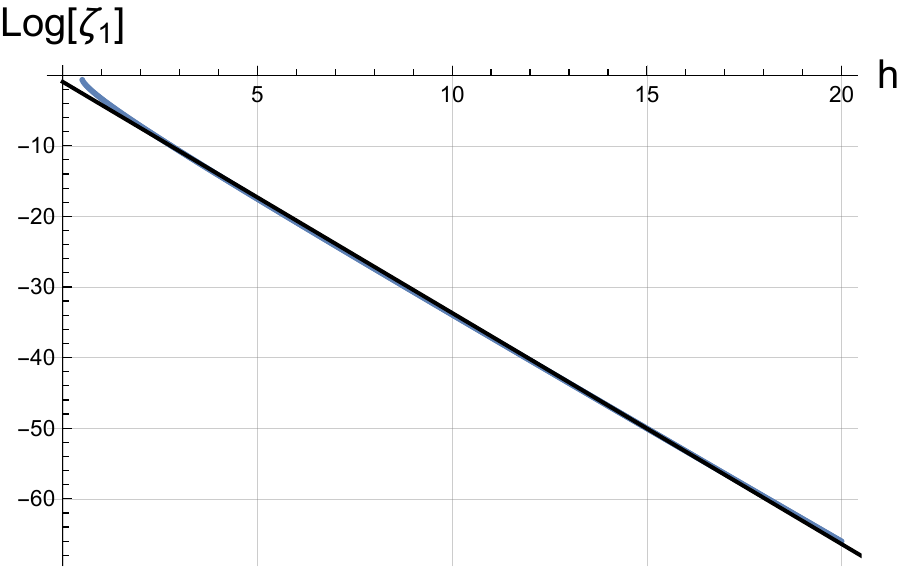}
\caption{$\zeta_1:= \mathfrak{E}_\sigma^{(h,1)}/c_\sigma$ for massive states with various values of $h.$ Left: We plot $\zeta_1$ for small $h.$ Right: Blue dots represent the numerical value of $\log(\zeta_1)$ at given $h.$ The black line is a linear fit for $\log(\zeta_1).$}\label{figure:2}
\end{figure}

With this in mind, we shall evaluate the contribution from a massive state with $(h,Q).$ We use the following expression to numerically evaluate the integral
\begin{equation}\label{eqn:massive contribution0}
\mathfrak{E}^{(h,Q)}_\sigma=c_\sigma \int_1^\infty \frac{dt}{2t^2} q^{h-\frac{1+Q}{4}}\vartheta_{1-Q,0}(2\tau)\,.
\end{equation}
For simplicity, we will only report the case of the annulus and the M{\"o}bius strip. For large $h$ we find
\begin{equation}
\mathfrak{E}^{(h,1)}\simeq 0.139 c_\sigma \exp(-3.271 h)\,,
\end{equation}
and
\begin{equation}
\mathfrak{E}^{(h,0)}\simeq  0.385 c_\sigma \exp(-3.272 h)\,.
\end{equation}
As expected, we find a very fast decay at large $h.$ We also report values of $\mathfrak{E}^{(h,Q)}$ at $h=|Q|/2$
\begin{equation}
\mathfrak{E}^{(1/2,1)}\simeq 0.509  c_\sigma\,,
\end{equation}
and
\begin{equation}
\mathfrak{E}^{(0,0)}\simeq 1.000 c_\sigma\,.
\end{equation}
For the numerical values of $\mathfrak{E}^{(h,Q)}$ for wide ranges of $h,$ see figure \ref{figure:1} and \ref{figure:2}.


On the other hand, one can consider a limit where at least one effective divisor in the Calabi-Yau orientifold shrinks. In such a case, one expects that infinitely many string states will become massless thereby inducing divergent correction to the K{\"a}hler potential by inspecting \eqref{eqn:massive contribution0}.\footnote{We thank Daniel Junghans, Gerben Venken, Arthur Hebecker, and Simon Schreyer for the related discussion.} As a result, one expects a correction to the Kahler potential that scales as
\begin{equation}
\frac{1}{T^a}\,,
\end{equation}
where $T$ is Einstein-frame divisor volume, and $a$ is a positive integer. This is in agreement with the conjecture proposed by Berg, Haack, and Pajer \cite{Berg:2007wt}. For related field theory analyses, see \cite{Cicoli:2007xp,Gao:2022uop}.


\subsection{Quintic threefold and conifold transitions}
We will now study how a particular class of small $\Bbb{P}^1$ cycles can affect the one-loop correction to the EH term. We will focus on the so-called \emph{nilpotent} $\Bbb{P}^1$ cycle, which we will denote by $\mathcal{C}$. For a detailed study of such curves, see for example \cite{Gendler:2022ztv}. The important property of such nilpotent $\Bbb{P}^1$ cycles is that one can shrink the volume of nilpotent $\Bbb{P}^1$ without shrinking the volume of any effective divisors. When $\mathcal{C}$ shrinks to a point, the Calabi-Yau geometry develops conifold singularities. Hence, at the level of $\mathcal{N}=2$ compactification, the point in the moduli space $t^*$ where $\text{Vol}(\mathcal{C})=0$ is a finite distance singularity and only a finite number of states become massless. We will first study the Quintic threefold and its conifold transition at the level of $\mathcal{N}=2$ compactification. Then, we will perform orientifolding of the Calabi-Yau threefold $X_3,$ which is obtained by the conifold transition from the Quintic threefold $Y_3.$ We will study the one-loop correction to the EH term and its corresponding correction to the K{\"a}hler potential when the nilpotent curve volume is extremely small.

Let us start with the GLSM description of $Y_3.$ There are five homogeneous coordinates $x_i,$ for $i=1,\dots,5,$ and their GLSM charge matrix is given as
\begin{equation}
\begin{array}{c|c|c|c|c}
x_1&x_2&x_3&x_4&x_5\\\hline
1&1&1&1&1
\end{array}
\end{equation}
Hence, the toric ambient variety $V_4$ is four-dimensional projective space $\Bbb{P}^4,$ and the Quintic threefold is defined as an anti-canonical hypersurface within $\Bbb{P}^4.$ The defining equation of $Y_3$ is a degree 5 polynomial in $x_i,$ whose form is
\begin{equation}
G= \sum_{i=0}^5 g^{(i)}(x_4,x_5)h^{(i)}(x_1,x_2,x_3)\,,
\end{equation}
where $g^{(i)}$ is a degree $i$ polynomial in $x_4$ and $x_5,$ and $h^{(i)}$ is a degree $5-i$ polynomial in $x_1,$ $x_2,$ and $x_3.$ We define an effective divisor class $[D]$ as a vanishing locus of $x_i,$ and define its dual curve class to be $[\mathcal{C}].$ By definition we have
\begin{equation}
\mathcal{C}\cdot D=1\,.
\end{equation}
We compute the Hodge vector of $D$
\begin{equation}
h^\bullet(D,\mathcal{O}_D)=(1,0,4)\,.
\end{equation}
We compute the intersection form $J$ of the Quintic threefold
\begin{equation}
J=\frac{5}{3!}t^3\,,
\end{equation}
where $t$ is volume of $\mathcal{C}.$ We denote the complexified volume of $\mathcal{C}$ by $z:=\int_{\mathcal{C}} (B_2+iJ).$The triple intersection number $K_{DDD}$ is given as
\begin{equation}
K_{DDD}=\partial_{t}^3J=5\,.
\end{equation}
We record a few genus-0 Gopakumar-Vafa (GV) invariants \cite{Gopakumar:1998ii,Gopakumar:1998jq} of the Quintic threefold\footnote{For a detailed study of the Quintic threefold and GV invariants therein, see \cite{Candelas:1990rm}.}
\begin{center}
\begin{tabular}{|c|c|c|c|c|c|}\hline
& $\mathcal{C}$&$2 \mathcal{C}$&$3\mathcal{C} $&$4\mathcal{C}$&$5\mathcal{C}$\\\hline
GV& 2875&609250&317206375&242467530000&229305888887625\\\hline
\end{tabular}
\end{center}

When we are at a generic point in the moduli space, the defining equation $G$ is regular at every point in the Calabi-Yau. Now, we tune complex structure moduli so that 
\begin{equation}
g^{(0)}(x_4,x_5)h^{(0)}(x_1,x_2,x_3)=0\,.
\end{equation}
Then, the defining equation $G$ can be written as
\begin{equation}
G=x_4 G_1 +x_5 G_2\,,
\end{equation}
where $G_1$ and $G_2$ are accordingly chosen degree 4 polynomials. We claim that this Quintic threefold has 16 conifolds singularities. To check this claim, we need to look for solutions to
\begin{equation}
G=\partial_i G=0\,,
\end{equation}
where $\partial_i G:=\partial G/\partial x_i.$ There are solutions to this set of equations
\begin{equation}
x_4=x_5=G_1=G_2=0\,.
\end{equation}
Because $G_1$ and $G_2$ are degree four polynomials, we find that there are point-like solutions where the Calabi-Yau $Y_3$ is singular. To resolve the singularities, we will blow up $Y_3$ such that $x_4=x_5=0$ becomes a subset of the Stanley-Reisner ideal. To do so, we introduce an exceptional divisor $E$ and its associated homogeneous coordinate $e,$ and rescale $x_4$ and $x_5$ as
\begin{equation}
x_4\mapsto e x_4\,,\qquad x_5\mapsto e x_5\,.
\end{equation}
After the introduction of the exceptional divisor, we have
\begin{equation}
G=\sum_{i=1}^{5} e^{i}g^{(i)}(x_4,x_5)h^{(i)}(x_1,x_2,x_3)\,.
\end{equation}
Because there is an overall factor of $e,$ we shall mod out this overall factor. As a result, we find the defining equation of $X_3$ is
\begin{equation}
\tilde{G}=\sum_{i=1}^{5} e^{i-1}g^{(i)}(x_4,x_5) h^{(i)}(x_1,x_2,x_3)\,.
\end{equation}
Note that the  corresponding GLSM for $X_3$ is given as
\begin{equation}
\begin{array}{c|c|c|c|c|c}
x_1&x_2&x_3&x_4&x_5&e\\\hline
1&1&1&0&0&1\\\hline
0&0&0&1&1&-1
\end{array}
\end{equation}
$X_3$ has $h^{1,1}(X_3)=2$ and $h^{2,1}(X_3)=86.$

Let us define divisor classes as
\begin{equation}
[D_1]:=\{ x_1=0\}\,,\qquad [D_2]:=\{x_4=0\}\,,\qquad [E]:=\{e=0\}\,.
\end{equation}
We pick a basis of the $H^2(X_3,\Bbb{Z})$ to be $\{ [D_1],[D_2]\}.$ We choose a dual basis of curves $\{ [\mathcal{C}_1], [\mathcal{C}_2]\}$ such that
\begin{equation}
\mathcal{C}_1\cdot D_1=1\,,\qquad \mathcal{C}_1\cdot D_2=0\,,\qquad \mathcal{C}_1\cdot E=1\,,
\end{equation}
and
\begin{equation}
\mathcal{C}_2\cdot D_1=0\,,\qquad \mathcal{C}_2\cdot D_2=1\,,\qquad \mathcal{C}_2\cdot E=-1\,.
\end{equation}
We will the denote volume of $\mathcal{C}_i$ as $t_i.$ Similarly, we define the complexified volume $z_i$ as 
\begin{equation}
z_i:=\int_{\mathcal{C}_i}\left(B_2+iJ\right)\,.
\end{equation}
We find the intersection form
\begin{equation}
J=\frac{5}{6} t_1^3+2t_1^2 t_2\,.
\end{equation}
Note that volumes of divisors $[D_1],$ $[D_2]$ and $[E]$ are given as
\begin{align}
\text{Vol}(D_1)=\frac{5}{2}t_1^2+2t_1t_2\,,\quad \text{Vol}(D_2)=2t_1^2\,,\quad \text{Vol}(E)=\frac{1}{2}t_1^2+4t_1t_2\,.
\end{align}
We compute the Hodge vectors
\begin{equation}
h^{\bullet}(D_1,\mathcal{O}_{D_1})=(1,0,4)\,,\quad h^\bullet(D_2,\mathcal{O}_{D_2})=(1,0,1)\,,\quad h^\bullet(E,\mathcal{O}_E)=(1,0,0)\,.
\end{equation} 
We compute genus-0 GV invariants of $X_3$\footnote{We used CYtools to compute the GV invariants \cite{Demirtas:2022hqf,cms}.}
\begin{center}
\begin{tabular}{|c|c|c|c|c|c|}\hline
&0& $\mathcal{C}_1$&$2 \mathcal{C}_1$&$3\mathcal{C}_1 $&$4\mathcal{C}_1$\\\hline
0&0&640&10032&288384&10979984\\\hline
$\mathcal{C}_2$&16&2144&231888&23953120&2388434784\\\hline
$2\mathcal{C}_2$&0&120&356368&144785584&36512550816\\\hline
$3\mathcal{C}_3$&0&-32&14608&144051072&115675981232\\\hline
$4\mathcal{C}_3$&0&3&-4920&5273880&85456640608\\\hline
\end{tabular}
\end{center}
It should be noted that the GV invariants of the Quintic can be obtained by summing over GV invariants of $X_3$ by
\begin{equation}
GV(n \mathcal{C})=\sum_{i=0}^\infty GV(n \mathcal{C}_1+i\mathcal{C}_2)\,,
\end{equation}
for $n\neq0.$ The curve class $\mathcal{C}_2$ is the nilpotent curve that we are looking for. Upon shirinking $\mathcal{C}_2,$ $X_3$ goes back to the singular $Y_3.$ We conclude that the Mori cone $\mathcal{M}(X_3)$ is generated by $[\mathcal{C}_1]$ and $[\mathcal{C}_2]$ over $\Bbb{Z}_{>0}.$ Similarly, the effective cone $\mathcal{E}(X_3)$ is generated by $[D_1],$ $[D_2]$ and $[E]$ over $\Bbb{Z}_{>0}.$ Note that in the limit where $\mathcal{C}_2$ shrinks, the divisor classes $[D_2]$ and $[E]$ recombine to yield 
\begin{equation}
[D_1]\equiv [D_2]+[E]\quad \text{for } t_2=0\,.
\end{equation}
It is important to stress that in $z_2\rightarrow0$ limit, all effective divisors have finite size and their volumes scale as $\mathcal{O}(t_1^2).$

As promised, we shall now perform an orientifolding to study $\mathcal{N}=1$ corrections to the EH action and the corresponding terms in $K^{(1)}$ when $\text{Vol}(\mathcal{C}_2)$ is small. One very important remark is in order before we delve into the orientifolding. At the level of $\mathcal{N}=2$ compactification, $\re{z}_i=\int_{\mathcal{C}_i}B_2 $ can take any real value$\mod\Bbb{Z}.$ But, after the orientifolding, $\int B_2$ can only take half-integral values and we shall carefully choose its value. With this in mind, we will proceed to find an O3/O7 orientifold of $X_3.$ We choose an orientifold action $\Omega$ that maps
\begin{equation}
x_5\mapsto -x_5\,.
\end{equation}
The orientifold $X_3/\Omega$ has two sets of O-planes: an O7-plane at $x_5=0,$ and an O3-plane at $x_1=x_2=x_3=0.$ We compute equivariant Hodge numbers of $X_3/\Omega$ using the results of \cite{Jefferson:2022ssj}
\begin{equation}
h^{1,1}_+=3\,,\quad h^{1,1}_-=0\,,\quad h^{2,1}_-=45\,,\quad h^{2,1}_+=41\,.
\end{equation}
As a result, we compute
\begin{equation}
\chi_f=18\,.
\end{equation}
To simplify the discussion, we will cancel the D7-brane tadpole locally by placing four D7-branes on the O7-plane at $x_5=0.$ In such a configuration, the D3-brane tadpole is given as
\begin{equation}
-Q_{D3}^{tadpole}=\frac{\chi_f}{4}=4+\frac{1}{2}\,.
\end{equation}
To saturate the D3-brane tadpole, we shall place four spacetime filling D3-branes at generic locations in $X_3/\Omega,$ and put one D3-brane on top of the O3-plane at $x_1=x_2=x_3=0.$ We will set $\re{z}_1=0$ and $\re{z}_2=1/2.$ The reason for choosing $\re z_2=1/2$ is to cancel the Freed-Witten anomaly of the seven-brane stack at $x_5=0$ \cite{Freed:1999vc}. An important comment is in order. Because of the half-integral B-flux $\re z_2=1/2,$ one cannot send $z_2\rightarrow 0$ limit even if the volume of $\mathcal{C}_2$ is extremely small. Therefore the orientifold $X_3/\Omega$ is now disconnected from an orientifold $Y_3/\overline{\Omega},$ where the orientifold action $\overline{\Omega}$ acts as $x_5\mapsto -x_5.$ But, the flop transition from $\im z_2>0$ to $\im z_2<0$ is not projected out by the orientifolding. Note that $Y_3/\overline{\Omega}$ has the equivariant Hodge numbers
\begin{equation}
h_+^{1,1}(Y_3)= 2\,,\quad h_-^{1,1}(Y_3)=0\,,\quad h^{2,1}_-(Y_3)=63\,,\quad h^{2,1}_+(Y_3)=38\,.
\end{equation}

Now, we will take $\im z_2=\epsilon$ to be very small and at the same time take $\im z_1$ to be large, so that the overall Calabi-Yau volume 
\begin{equation}
\mathcal{V}=\frac{5}{6} t_1^3+2t_1^2 t_2\,,
\end{equation}
is also large. 

We first study the Klein bottle contribution. Had the $\re z_2=0$ been zero, $h_-^{2,1}(Y_3)-h_-^{2,1}(X_3)$ many of the frozen complex structure moduli of $Y_3/\overline{\Omega}$ would have become massless and moduli again in $z_2\rightarrow 0$ limit. Similarly, in such a limit, $h^{2,1}(X_3)-h^{2,1}(Y_3)$ many vector multiplets can be deformed to be massive. That being said, in a small $\epsilon$ limit, $\mathcal{O}(10)$ irreducible states of the closed string theory should become almost massless. Because the lowest lying modes dominate $\delta E,$ we can use those almost massless irreducible states to estimate the size of $\delta E_\mathcal{K}.$ We denote the contributions from the almost massless irreducible states by $\delta E_{K}^{\epsilon},$ which is estimated to be
\begin{equation}
\delta E_{\mathcal{K}}^{\epsilon}= 2^{-5} \pi^{-2} \times \mathcal{O}(10)= \mathcal{O}(10^{-2})\,.
\end{equation}
Now, let us study the contributions from open strings extended from and to the seven-brane stack. Carefully counting the change in the number of moduli is outside the scope of this paper. Instead, we will present a rough estimate as the order of magnitude is what matters here. The Hodge vector of the divisor would have changed from $(1,0,1)$ to $(1,0,4).$ So, in the upstairs picture, one can see that a single seven-brane wrapped on the divisor $D_5$ would gain three deformation moduli. Note that the Hodge vector of the divisor class $[2D_5]$ changes from $(1,0,2)$ to $(1,0,14).$ We also compute that the equivariant Hodge vectors would change from \begin{equation}
h_+^\bullet(2D_5,\mathcal{O}_{2D_5})=(1,0,1)\,,  \quad h_-^\bullet(2D_5,\mathcal{O}_{2D_5})=(0,0,1)\,,
\end{equation}
to  
\begin{equation}
h_+^\bullet(2D_5,\mathcal{O}_{2D_5})=(1,0,4)\,,  \quad h_-^\bullet(2D_5,\mathcal{O}_{2D_5})=(0,0,10)\,.
\end{equation}
As a single seven-brane in the downstairs picture corresponds to two seven-branes wrapped on the same divisor, for a single seven-brane in the downstairs picture, we estimate that there should be $\mathcal{O}(10)$ states becoming almost massless in small $\epsilon$ limit. Therefore, we estimate
\begin{equation}
\frac{1}{2}\left(\delta E_{A,D7-D7}^{\epsilon}+\delta E_{\mathcal{M},D7-D7}^{\epsilon}\right)=\mathcal{O}(10^{-3})\,.
\end{equation}
Because the small $\mathcal{C}_2$ volume does not induce any new zero modes for D3-D7 and D3-D3 strings, we don't expect a significant contribution. Therefore, we conclude that the change in $\delta E$ when shrinking volume of $\mathcal{C}_2$ while fixing $\int_{\mathcal{C}_2}B_2=1/2$ is estimated to be
\begin{equation}
\delta E^{\epsilon}-\delta E^{generic} =\mathcal{O}(10^{-2})\,.
\end{equation}

Now, let us finally study its effect on the one-loop corrected K{\"a}hler potential $K_{\phi\bar{\phi}}^{(1)}.$ The one-loop correction to the EH action contribution is
\begin{equation}
e^{2\Phi_4}\frac{K_{\phi\bar{\phi},\delta E}^{(1)}}{K_{\phi\bar{\phi}}^{(0)}}= \mathcal{O}\left( e^{2\Phi_4} \delta E\right)\,.
\end{equation}
Therefore, we find that under the limit $\im z_2\rightarrow 0,$ we have
\begin{equation}
\delta\left(e^{2\Phi_4}\frac{K_{\phi\bar{\phi},\delta E}^{(1)}}{K_{\phi\bar{\phi}}^{(0)}}\right) = \mathcal{O}\left(10^{-3} \frac{g_s^2}{\mathcal{V}}\right)\,.
\end{equation}
Quite happily, when the overall Calabi-Yau volume is large and string coupling is small, the one-loop correction to the K{\"a}hler potential due to the one-loop correction to the EH term is very small.

\section{Discussion and future directions}\label{sec:discussion}
In this note, we studied the string one-loop corrections to the Einstein-Hilbert action in string frame. Because these corrections are required to fully determine the string one-loop corrected K{\"a}hler potential of moduli fields, it is very important to compute the loop corrections to the EH term. 

One surprising result is that the outcome of the computation is that the string one loop correction is at large determined by the new supersymmetric index \cite{Cecotti:1992qh}
\begin{equation}
\text{Tr}_R \left( (-1)^F F q^{L_0-\frac{3}{8}}\right)\,,
\end{equation}
for example recall the annulus contribution
\begin{equation}
\delta E_A=\frac{1}{2^9\pi^2}\int_0^\infty \frac{dt}{2t^2} \Bigl[\text{Tr}_R \left((-1)^{F-\frac{3}{2}}F q^{L_0-\frac{3}{8}}\right)^{open}_{int}+\frac{3}{2} \left(n^+_A-n^-_A\right)\Bigr]\,.
\end{equation}
This is a very nice feature of the one-loop correction for the following reason. As was studied in \cite{Cecotti:1992qh}, the new supersymmetric index is only sensitive to the F-terms of the internal theory or the chiral rings of the internal CFT. This is a much more favorable situation than a generic amplitude which may in general depend on D-terms of the internal theory as well, which are much more difficult to understand. 

The claim that the moduli dependence of $\delta E$ is determined by the new supersymmetric index seems quite radical because the new supersymmetric index typically shows up in topological string amplitudes that compute the F-terms of the low energy supergravity of \emph{physical} superstring theories \cite{Bershadsky:1993cx}. Therefore the computations performed in \S\ref{sec:graviton scattering at one loop} may look very surprising to some cautious readers. In light of this, we shall provide an intuitive explanation of the origin of the new supersymmetric index. To do so, it is useful to study the holomorphic gauge coupling of gauge fields living on D-branes in 4d $\mathcal{N}=1$ effective supergravity obtained from type II string compactification on Calabi-Yau orientifolds. As is well known, the kinetic term of a gauge field is given by the F-term
\begin{equation}
\int d^2 \theta f \mathcal{W}_\alpha \mathcal{W}^\alpha\,,
\end{equation}
where $\mathcal{W}$ is the field strength superfield for the vector multiplet, and $f$ is the holomorphic gauge coupling. In \cite{Bershadsky:1993cx} it was conjectured that the threshold correction to the \emph{effective} gauge coupling is computed by open topological string theory partition function at one-loop
\begin{equation}
\int \frac{dt}{2t} \text{Tr}_R \left((-1)^{F-\frac{3}{2}}F q^{L_0-\frac{3}{8}}\right)_{int}\,,
\end{equation}
which is shown to satisfy the holomorphic anomaly equation \cite{Bershadsky:1993ta,Bershadsky:1993cx,Walcher:2007tp,Walcher:2007qp}. Partial proofs of the open string version of the BCOV conjecture were given in \cite{Antoniadis:2005sd,Kim:2023cbh}.

An efficient way to compute the threshold correction to the effective gauge coupling is to use the background field method. One can start by computing the string one-loop partition function $\Lambda(\mathcal{F}),$ summed over the even spin structure, where the non-trivial gauge flux $\mathcal{F}$ of a D-brane gauge theory is turned on. Then, the threshold correction to the gauge coupling can be read off by computing \cite{Antoniadis:1999ge,Berg:2004ek,Conlon:2009xf,Alexandrov:2022mmy}
\begin{equation}
\frac{\partial^2}{\partial \mathcal{F}^2}\Lambda(\mathcal{F})|_{\mathcal{F}=0}\,.
\end{equation}
Because the non-trivial $\mathcal{F}$ corresponds to the twisted boundary conditions for open strings, taking the double derivatives in $\mathcal{F}$ results in the inclusion of the factor \cite{Alexandrov:2022mmy}
\begin{equation}
t^2\frac{\vartheta_{\alpha,\beta}''(0|\tau)}{\vartheta_{\alpha,\beta}(0|\tau)}
\end{equation}
in the partition function of the open string. Note that $(\alpha,\beta)$ denotes the spin structure of the open string worldsheet. Because the one-loop partition function for open strings with the spin structure $(\alpha,\beta)$ in the absence of the worldvolume flux is given by
\begin{equation}\label{eqn:partition 1}
c\int \frac{dt}{t^3} (-1)^{\alpha+\beta}\frac{\vartheta_{\alpha,\beta}(0|\tau)}{\eta(\tau)^3} \text{Tr}_{\alpha}\left( (-1)^{\beta F} q^{L_0-\frac{3}{8}}\right)_{int}\,,
\end{equation}
where $c$ is a numerical factor and the trace is over the states in the internal CFT, one can expect that the one-loop amplitude of the form
\begin{equation}\label{eqn:topological string 1}
\sum_{\alpha,\beta \text{ even}}\int \frac{dt}{t} (-1)^{\alpha+\beta} \frac{\vartheta_{\alpha,\beta}''(0|\tau)}{\eta(\tau)^3} \text{Tr}_{\alpha}\left( (-1)^{\beta F} q^{L_0-\frac{3}{8}}\right)_{int}
\end{equation}
must be related to the open topological string theory amplitude at one loop as was shown in \cite{Kim:2023cbh}.  

Now let us compare \eqref{eqn:topological string 1} to \eqref{eqn:delta E intermediate}
\begin{equation}\label{eqn:delta E intermediate2}
\delta E_A=-\frac{1}{2^{12}\pi^4}\sum_{\alpha,\beta \text{ even}} \int_0^\infty \frac{dt}{t^2} (-1)^{\alpha+\beta} \frac{\vartheta_{\alpha,\beta}''(0|\tau)}{\eta(\tau)^3}\text{Tr}_{\alpha}\left((-1)^{\beta F} q^{L_0-\frac{3}{8}} \right)^{open}_{int} \,.
\end{equation}
The integrand of  \eqref{eqn:topological string 1} is $t$ times the integrand of \eqref{eqn:delta E intermediate2} modulo the numerical factors! Hence,  we can conclude from this comparison that the Einstein-Hilbert action at string one-loop is indeed determined by the new supersymmetric index and the number of (anti) BPS states. 

The reason why we have strikingly similar structures in \eqref{eqn:topological string 1} and \eqref{eqn:delta E intermediate2} is quite simple. An equivalent way to compute the threshold correction \eqref{eqn:delta E intermediate2} is to insert two vertex operators for the gauge field. Because the open string vertex operator only contains either holomorphic or anti-holomorphic fields, each open string vertex operator can at the most contain two free fermions in $(0,0)$ picture. Therefore, the only non-vanishing contribution to the threshold correction is essentially coming from four fermion correlators, after the spin sum. Note that the very same structure is found in \eqref{eqn:delta E intermediate} for a different reason. In the graviton two point function computation, we concluded that the four fermion contractions are the only non-trivial contributions. But, these contractions come with the bosonic contractions as well. What we found in \eqref{eqn:delta E intermediate0} was that, after integrating over the vertex position moduli, the bosonic contractions drop out and all that remains is essentially the four fermion contraction. Therefore, we found the similar structure in the graviton two point function.

Although we have not evaluated the one-loop corrected Einstein-Hilbert action in explicit models, we can still learn general lessons about when the one-loop correction gets dangerously large. For example, we concluded that the correction to the Einstein-Hilbert action is divergent if infinitely many states become massless. As an application, in \S\ref{sec:small cycles}, we estimated the size of the one-loop correction to the EH action in the limit where nilpotent curves are small. The soothing conclusion we found is that its effect on the one-loop corrected K{\"a}hler potential is minuscule when the overall Calabi-Yau volume is large and string coupling is small. We hope that our result can be used to learn more interesting lessons about type II compactifications on Calabi-Yau orientifolds.

Let us now discuss possible future directions. 
\begin{itemize}
\item In the case of a particular moduli  integral of the new supersymmetric index
\begin{equation}
Z=\int \frac{dt}{2t} \text{Tr}_R\left((-1)^F F q^{L_0-\frac{3}{8}}\right)_{int}\,,
\end{equation}
one can understand the holomorphic anomalies associated with the moduli derivatives of Z exactly \cite{Bershadsky:1993cx,Bershadsky:1993ta,Walcher:2007tp,Walcher:2007qp}. It might be possible to derive a similar holomorphic anomaly equation for the one-loop correction to the Einstein-Hilbert term as well.
\item Direct computation of the new supersymmetric index for open strings in Calabi-Yau orientifold compactifications is not very well understood. It is very important to make progress on the computation of such indices.
\item It is a reasonable expectation that the string one-loop correction to the moduli kinetic term is of the same order as the string one-loop correction to the EH action as both of them arise from the same diagram, and can be computed by inserting two vertex operators and reading off the kinematic factors. Although very plausible, it is nevertheless very crucial to check this claim via explicit computations. Unfortunately, the string one-loop correction to the moduli kinetic term in string-frame has yet been computed only in toroidal orientifold compactifications \cite{Berg:2005ja,Berg:2014ama}. To sum up, it would be extremely important to develop tools to compute such a correction in more generic compactifications. 
\item In this work, we saturated the D-brane tadpole with spacetime filling D-branes. In more realistic compactifications, NSNS and RR fluxes are ubiquitous. Therefore, it is important to develop methods to compute the string loop corrections in flux backgrounds.
\end{itemize}
\section*{Acknowledgements}
The work of MK was supported by a Pappalardo fellowship. We thank the referee for valuable suggestions that improved the draft significantly. MK thanks Atakan Hilmi F{\i}rat, Liam McAllister, Mehmet Demirtas, Wati Taylor, Daniel Junghans, Gerben Venken, Arthur Hebecker, Harold Erbin, Xi Yin, Ying-Hsuan Lin, and Simon Schreyer for the discussions. MK thanks Patrick Jefferson, Michael Haack, and Marcus Berg for comments on the draft. MK thanks Daniel Harlow and Hirosi Ooguri for their encouragement. 
\appendix
\newpage
\section{Jacobi theta function}\label{app:jacobi theta}
In this section, we summarize the convention for Jacobi theta functions and many useful identities. We will mostly follow the conventions of \cite{Alexandrov:2022mmy}. For $\alpha,\beta=0,1$ we define
\begin{equation}
\vartheta_{\alpha,\beta}(z|\tau):=\sum_{n\in \Bbb{Z}+\frac{\alpha}{2}}e^{i\pi n \beta}q^{n^2/2}y^n\,,\qquad q=e^{2\pi i\tau}\,,\qquad y=e^{2\pi iz}\,.
\end{equation}
We also define 
\begin{equation}
\vartheta_{\alpha,\beta}(\tau):=\vartheta_{\alpha,\beta}(0|\tau)\,.
\end{equation}
We write $\vartheta_{\alpha,\beta}(z|\tau)$ for $(\alpha,\beta)=(0,0),~(0,1),~(1,0),~(1,1)$
\begin{align}
\vartheta_{0,0}(z|\tau)=&\prod_{n=1}^{\infty}(1-q^n) \left(1+(y+y^{-1})q^{n-\frac{1}{2}}+q^{2n-1}\right)\,,\\
\vartheta_{0,1}(z|\tau)=&\prod_{n=1}^{\infty}(1-q^n)\left(1-(y+y^{-1})q^{n-\frac{1}{2}}+q^{2n-1}\right)\,,\\
\vartheta_{1,0}(z|\tau)=&q^{\frac{1}{8}}(y^{\frac{1}{2}}+y^{-\frac{1}{2}})\prod_{n=1}^\infty (1-q^n)\left(1+(y+y^{-1})q^n+q^{2n}\right)\,,\\
\vartheta_{1,1}(z|\tau)=&iq^{\frac{1}{8}}(y^{\frac{1}{2}}-y^{-\frac{1}{2}})\prod_{n=1}^\infty (1-q^n)\left(1-(y+y^{-1})q^n+q^{2n}\right)\,.
\end{align}
The Jacobi theta functions enjoy quasi-periodicity, 
\begin{align}
\vartheta_{\alpha,\beta}\left(z+\frac{1}{2}|\tau\right)=&\vartheta_{\alpha,\beta+1}(z|\tau)\,,\\
\vartheta_{\alpha,\beta}\left(z+\frac{\tau}{2}|\tau\right)=&e^{-i\pi \beta/2}q^{-\frac{1}{8}}y^{-\frac{1}{2}}\vartheta_{\alpha+1,\beta}(z|\tau)\,,\\
\vartheta_{\alpha+2,\beta}(z|\tau)=&\vartheta_{\alpha,\beta}(z|\tau)\,,\\
\vartheta_{\alpha,\beta+2}(z|\tau)=&e^{i\alpha\pi}\vartheta_{\alpha,\beta}(z|\tau)\,.
\end{align}
The Jacobi theta functions satisfy the Jacobi identity
\begin{equation}
\vartheta_{0,0}(\tau)^4=\vartheta_{1,0}(\tau)^4+\vartheta_{0,1}(\tau)^4\,.
\end{equation}

We define the Dedekind eta function as
\begin{equation}
\eta(\tau):= q^{\frac{1}{24}} \prod_{n=1}^\infty (1-q^n)\,.
\end{equation}
We record useful identities involving the Dedekind eta function and the Jacobi theta functions
\begin{align}
\partial_z \vartheta_{1,1}(z|\tau)|_{z=0}=-2\pi \eta(\tau)^3\,,
\end{align}
\begin{equation}
\vartheta_{0,0}(\tau)=\frac{\eta(\frac{1}{2}(\tau+1))}{\eta(\tau+1)}=\frac{\eta(\tau)^5}{\eta(\frac{1}{2}\tau)^2\eta(2\tau)^2}\,,\quad \vartheta_{0,1}=\frac{\eta(\frac{1}{2}\tau)^2}{\eta(\tau)}\,\quad \vartheta_{1,0}(\tau)=\frac{2\eta(2\tau)^2}{\eta(\tau)}\,,
\end{equation}
and
\begin{equation}
\vartheta_{0,0}(\tau)\vartheta_{1,0}(\tau)\vartheta_{0,1}(\tau)=2\eta(\tau)^3\,.
\end{equation}

We write an important identity \cite{Epple:2004ra}
\begin{equation}\label{eqn:identity imp1}
\left(\frac{\vartheta_{s_1,s_2}(z|\tau)\vartheta_{1,1}'(0|\tau)}{\vartheta_{s_1,s_2}(0|\tau)\vartheta_{1,1}(z|\tau)}\right)^2=\frac{\vartheta_{s_1,s_2}''(0|\tau)}{\vartheta_{s_1,s_2}(0|\tau)}-\partial_z^2\log\vartheta_{1,1}(z|\tau)\,.
\end{equation}

The Jacobi theta functions obey addition rules \cite{gradshteyn2014table}
\begin{align}\label{eqn:addition rule1}
\vartheta_{0,0}(z_1|\tau)\vartheta_{0,0}(z_2|\tau)=&\vartheta_{0,0}(z_1+z_2|2\tau)\vartheta_{0,0}(z_1-z_2|2\tau)+\vartheta_{1,0}(z_1+z_2|2\tau)\vartheta_{1,0}(z_1-z_2|2\tau)\,,\\\label{eqn:addition rule2}
\vartheta_{0,0}(z_1|\tau)\vartheta_{0,1}(z_2|\tau)=&\vartheta_{0,1}(z_1+z_2|2\tau)\vartheta_{0,1}(z_1-z_2|2\tau)-\vartheta_{1,1}(z_1+z_2|2\tau)\vartheta_{1,1}(z_1-z_2|2\tau)\,,\\\label{eqn:addition rule3}
\vartheta_{0,1}(z_1|\tau)\vartheta_{0,1}(z_2|\tau)=&\vartheta_{0,0}(z_1+z_2|2\tau)\vartheta_{0,0}(z_1-z_2|2\tau)-\vartheta_{1,0}(z_1+z_2|2\tau)\vartheta_{1,0}(z_1-z_2|2\tau)\,,\\\label{eqn:addition rule4}
\vartheta_{1,0}(z_1|\tau)\vartheta_{1,0}(z_2|\tau)=&\vartheta_{1,0}(z_1+z_2|2\tau)\vartheta_{0,0}(z_1-z_2|2\tau)+\vartheta_{0,0}(z_1+z_2|2\tau)\vartheta_{1,0}(z_1-z_2|2\tau)\,,\\\label{eqn:addition rule5}
\vartheta_{1,1}(z_1|\tau)\vartheta_{1,1}(z_2|\tau)=&\vartheta_{0,0}(z_1+z_2|2\tau)\vartheta_{1,0}(z_1-z_2|2\tau)-\vartheta_{1,0}(z_1+z_2|2\tau)\vartheta_{0,0}(z_1-z_2|2\tau)\,.
\end{align}

For later use, we define
\begin{equation}
f_{k,Q}(z,\tau):= \frac{1}{\eta(\tau)}e^{i\pi \tau Q^2/k}e^{2\pi i Qz}\vartheta_{0,0}(kz+Q\tau|k\tau)\,.
\end{equation}
We write useful identities involving $f_{k,Q}(z,\tau)$
\begin{equation}
f_{1,0}(z,\tau)=\frac{\vartheta_{0,0}(z|\tau)}{\eta(\tau)}\,,\quad f_{2,0}(z,\tau)=\frac{\vartheta_{0,0}(2z|2\tau)}{\eta(\tau)}\,,\quad f_{2,1}(z,\tau)=\frac{\vartheta_{1,0}(2z|2\tau)}{\eta(\tau)}\,,
\end{equation}
and
\begin{equation}
f_{3,1}(z,\tau)-f_{3,-1}(z,\tau)=0\,,
\end{equation}
for $z=0,1/2,\tau/2,$ and
\begin{equation}
q^{\frac{3}{8}}\left(f_{3,1}(z,\tau)-f_{3,-1}(z,\tau)\right)=2\,,
\end{equation}
for $z=(1+\tau)/2.$
\section{Extended $\mathcal{N}=2$ superconformal algebra and its representation theory.}\label{app:N=2 superconformal algebra}
In this section, we summarize the representation theory of the extended $\mathcal{N}=2$ superconformal algebra \cite{Eguchi:1988vra,Odake:1988bh,Odake:1989ev,Odake:1989dm}. 

Let us first collect the OPEs of $\mathcal{N}=2$ superconformal generators: the energy momentum tensor $T,$ super currents $G$ and $\tilde{G},$ and the $U(1)_R$ current $I,$ 
\begin{align}
T(z)T(w)=&\frac{c}{2(z-w)^4}+\frac{2T(w)}{(z-w)^2}+\frac{\partial T(w)}{z-w}+\dots\,,\\
I(z)I(w)=&\frac{c}{3(z-w)^2}+\dots\,,\\
I(z)G(w)=&\frac{1}{z-w}G(w)+\dots\,,\\
I(z)\tilde{G}(w)=&-\frac{1}{z-w}\tilde{G}(w)+\dots\,,\\
G(z)\tilde{G}(w)=&\frac{2c}{3(z-w)^3}+\frac{2I(w)}{(z-w)^2}+\frac{1}{z-w}(\partial I(w)+2T(w))+\dots\,,\\
G(z)G(w)=&\text{regular}\,,\\
\tilde{G}(z)\tilde{G}(w)=&\text{regular}\,.
\end{align}
The extended $\mathcal{N}=2$ superconformal algebra is obtained by adding the spectral flow generators $X$ and $\tilde{X},$ and their superpartners $Y$ and $\tilde{Y}.$  

We shall summarize the character formulas for the extended superconformal algebra. For an irreducible representation $r,$ we define the character as
\begin{equation}
ch_\bullet^{(r)}(z,\tau):=\text{Tr}_{\bullet,r}\left(q^{L_0-\frac{3}{8}}y^{I_0}\right)\,,
\end{equation}
where $\bullet$ can be Neveu-Schwartz (NS) sector or Ramond (R) sector. Similarly, we define the character formulas with the GSO projection as
\begin{equation}
ch_{\tilde{NS}}^{(r)}(z,\tau):= \text{Tr}_{NS,r}\left((-1)^{I_0}q^{L_0-\frac{3}{8}}y^{I_0}\right)\,,
\end{equation}
and
\begin{equation}
ch_{\tilde{R}}^{(r)}(z,\tau):=\text{Tr}_{R,r}\left((-1)^{I_0-\frac{3}{2}}q^{L_0-\frac{3}{8}}y^{I_0}\right)\,.
\end{equation}
Note that for the character $ch_{\tilde{R}}^{(r)},$ we included the factor $(-1)^{-3/2},$ because $U(1)_R$ charge of states in the R sector is fractional and integral concerning the $U(1)_R$ charge of the vacuum state, whose charge is $3/2.$ The character formulas satisfy the following relations
\begin{align}
ch_{\tilde{NS}}^{(r)}(z,\tau)=&ch_{NS}^{(r)}(z+1/2,\tau)\,,\\
ch_{R}^{(r)}(z,\tau)=&q^{\frac{3}{8}}y^{\frac{3}{2}}ch_{NS}^{(r)}(z+\tau/2,\tau)\,,\\
ch_{\tilde{R}}^{(r)}(z,\tau)=&e^{-3 i\pi/2}ch_R^{(r)}(z+1/2,\tau)\,.
\end{align}
Frequently, we will use the following notations
\begin{align}\label{eqn:characters}
&g_{00}^{(r)}:=\text{Tr}_{NS,r}\left( q^{L_0-\frac{3}{8}} \right)\,,\quad g_{01}^{(r)}:=\text{Tr}_{NS,r}\left( q^{L_0-\frac{3}{8}} (-1)^{I_0}\right)\,,\\
&g_{10}^{(r)}:=\text{Tr}_{R,r}\left( q^{L_0-\frac{3}{8}} \right)\,,\quad \,\,\,g_{11}^{(r)}:=\text{Tr}_{R,r}\left( q^{L_0-\frac{3}{8}} (-1)^{I_0-\frac{3}{2}}\right)\,.
\end{align}
Note that following \cite{Lerche:1989uy,Odake:1989ev,Antoniadis:1992rq,Antoniadis:1992sa,Kaplunovsky:1995jw,Kim:2023cbh}, we will identify $I_0\equiv F,$ where $F$ is the fermion number of states/operators on the worldsheet cft. 

Let us now summarize the character formulas for irreducible representations. We shall start with massive representations which satisfy $h>|Q|/2.$ As a state in the R-sector can be obtained by a half-integral spectral flow from a state in the NS sector, we will label irreducible states with $(h,Q)$ of the highest weight state of the corresponding irreducible state in the NS sector. In the NS sector, the $U(1)_R$ charge of the highest weight state can take values from $-1,0,1.$ Because the character for $-Q$ is the same as the character for $Q,$ we shall only consider the character formulas for $Q>0.$ The character formula for all sectors takes the form
\begin{equation}
g_{\alpha\beta}^{(h,Q)}:= q^{\frac{3\alpha}{8}}g\left(\frac{\alpha\tau+\beta}{2},\tau;h,Q\right)\,,
\end{equation}
where we define
\begin{equation}
g(z,\tau;h,Q):= \frac{q^{h-\frac{1+Q^2}{4}}}{\eta(\tau)} f_{1,0}(z,\tau)f_{2,Q}(z,\tau)\,.
\end{equation}
We record useful identities
\begin{equation}
g_{\alpha\beta}^{(h,Q)}=e^{-i\pi \alpha\beta/2} (-1)^{\beta Q}\frac{q^{h-\frac{1+Q}{4}}}{\eta(\tau)^3}\vartheta_{\alpha,\beta}(\tau)\vartheta_{\alpha+Q,0}(2\tau)\,,
\end{equation}
and
\begin{equation}\label{eqn:the new supersymmetric index id}
\frac{1}{2\pi i}\partial_z ch_{\tilde{R}}^{(h,Q)}(z,\tau)|_{z=0}=\frac{1}{2\pi i}\partial_z q^{\frac{3}{8}}g(z,\tau;h,Q)|_{z=(1+\tau)/2}=(-1)^Q q^{h-\frac{1+Q}{4}}\vartheta_{1-Q,0}(2\tau)\,,
\end{equation}
for $h\geq|Q|/2.$

Now, let us study massless representations. There are three different states: the vacuum state with $(h,Q)=0,$ and $(\pm)$ states with $(h,Q)=(1/2,\pm 1).$ The character for the massless representations are obtained by replacing $g(z,\tau;h,Q)$ with
\begin{align}\label{eqn:massless characters}
g^{(vac)}(z,\tau)=& g(z,\tau;0,0)-g\left(z,\tau;\frac{1}{2},1\right)\,,\\
g^{(\pm)}(z,\tau)=&\pm\frac{1}{2}(f_{3,1}(z,\tau)-f_{3,-1}(z,\tau))+\frac{1}{2}g\left(z,\tau;\frac{1}{2},1\right)\,.\label{eqn:massless characters2}
\end{align}
Note that
\begin{equation}
f_{3,Q}\left(z+\frac{1+\tau}{2}\biggr|\tau\right)=-\frac{i}{\eta(\tau)}(-1)^Qq^{\frac{Q^2}{6}-\frac{3}{8}}e^{\pi i (2Q-3) z}\vartheta_{1,1}(3z+Q\tau|3\tau)\,,
\end{equation}
and $\vartheta_{1,1}(z|3\tau)$ is an odd function in $z.$ We compute
\begin{equation}
\frac{1}{2\pi i}q^{\frac{3}{8}} \partial_z \left(f_{3,Q}(z,\tau)\right)|_{z=(1+\tau)/2}=\frac{2Q-3}{2}q^{\frac{3}{8}}  f_{3,Q}\left(\left.\frac{1+\tau}{2}\right|\tau\right)+\frac{3(-1)^{Q+1}}{2\pi\eta(\tau)}q^{\frac{Q^2}{6}}\vartheta_{1,1}'(Q\tau|3\tau)\,.
\end{equation}
As a result, it follows that
\begin{equation}\label{eqn:massless f}
\frac{1}{2\pi i}q^{\frac{3}{8}}\partial_z (f_{3,1}(z,\tau)-f_{3,-1}(z,\tau))|_{z=(1+\tau)/2}=-3\,.
\end{equation}
\section{Green's function manipulations}\label{app:Greens function}
In this section, we compute 
\begin{equation}\label{eqn app:graviton two point}
K(p_1,p_2,\epsilon_1,\epsilon_2)=\langle V_g^{(0,0)}(z_1,p_1,\epsilon_1)V_g^{(0,0)}(z_2,p_2,\epsilon_2)\rangle_\sigma
\end{equation}
for an arbitrary Riemann surface $\sigma$ of genus 1 with or without boundaries. Let us recall that the graviton vertex operator in the $(0,0)$ picture is given as
\begin{equation}
V_g^{(0,0)}(z,p,\epsilon)=-\frac{2g_c}{\alpha'}\epsilon_{\mu\nu} \left(i\partial X^\mu +\frac{\alpha'}{2}p \cdot \psi \psi^\mu\right)\left(i\bar{\partial}X^\nu + \frac{\alpha'}{2} p\cdot \bar{\psi}\bar{\psi}^\nu\right) e^{i p\cdot X}\,.
\end{equation} 
We shall impose the incomplete on-shell condition
\begin{equation}
p_1^2=p_2^2=p_1\cdot p_2=p_{1\mu}\epsilon_{1}^{\mu\nu}=p_{2\mu}\epsilon_2^{\mu\nu}=\eta_{\mu\nu}\epsilon^{\mu\nu}_1=\eta_{\mu\nu}\epsilon^{\mu\nu}_2=0\,,
\end{equation}
to simplify the evaluation of \eqref{eqn app:graviton two point}.

We write
\begin{align}
K= &\frac{4g_c^2\epsilon_{1\mu\nu}\epsilon_{2\rho\sigma}}{\alpha'^2}\left\langle e^{i p_1\cdot  X_1} e^{ip_2\cdot X_2}\left(i\partial X_1^\mu +\frac{\alpha'}{2} p_1\cdot \psi_1 \psi_1^\mu\right)\left(i\bar{\partial} X_1^\nu +\frac{\alpha'}{2} p_1\cdot \bar{\psi}_1 \bar{\psi}_1^\nu\right)\right.\nonumber\\
&\qquad\qquad\qquad\left.\times\left(i\partial X_2^\rho +\frac{\alpha'}{2} p_2\cdot \psi_2 \psi_2^\rho\right)\left(i\bar{\partial} X_2^\sigma +\frac{\alpha'}{2} p_2\cdot \bar{\psi}_2 \bar{\psi}_2^\sigma\right)\right\rangle \,.
\end{align}
A few comments are in order. We are attempting to find terms of order $\mathcal{O}(p^2).$ Therefore, we need to contract at least four fermions. One might be worried that we also need to consider eight fermion contractions because it is possible that there can be a pole when vertex operators are colliding and when such a pole is present, eight fermion contractions can in principle contribute to order $\mathcal{O}(p^2)$ terms. But, it should be noted that such a pole always comes with $p_1\cdot p_2$ factor we are setting to zero, hence such a pole cannot exist. With this understanding, we will from now on ignore $e^{ip_1\cdot X_1}e^{ip_2\cdot X_2}$ factor and focus on four fermion contractions.

In total, there are four different ways to contract the bosonic part: $\langle\partial X_1^\mu \partial X_2^\rho\rangle,$ $\langle \partial X_1^\mu \bar{\partial} X_2^\sigma\rangle,$ $\langle \bar{\partial}X_1^\nu\partial X_2^\rho\rangle,$ and $\langle \bar{\partial}X_1^\nu\bar{\partial}X_2^\sigma\rangle.$ We will study the case $\langle \partial X_1^\mu\partial X_2^\rho\rangle$ in detail, and spell out the results for the other cases. After contracting $\langle \partial X_1^\mu \partial X_2^\rho\rangle,$ we have to contract the left fermions 
\begin{equation}
\left\langle\frac{\alpha'^2}{4} p_1\cdot \bar{\psi}_1\bar{\psi}_1^\nu p_2\cdot \bar{\psi}_2\bar{\psi}_2^\sigma\right\rangle\,.
\end{equation}
As we are imposing the incomplete on-shell condition, which includes the condition $p_1\cdot p_2=0,$ contracting $p_1\cdot\bar{\psi}_1$ and $p_2\cdot \bar{\psi}_2$ will produce zero. Therefore, the only non-zero contribution is by the following contraction
\begin{equation}
\frac{\alpha'^2}{4} \left\langle p_1\cdot \bar{\psi}_1\bar{\psi}_2^\sigma \right\rangle \times \left\langle \bar{\psi}_1^\nu p_2\cdot \bar{\psi}_2\right\rangle\,,
\end{equation}
which equals to
\begin{equation}
\frac{\alpha'^2}{4} p_{1\alpha}p_{2\gamma}\langle \bar{\psi}_2^\sigma\bar{\psi}_1^{\alpha}\rangle \langle \bar{\psi}_2^\gamma\bar{\psi}_1^\nu\rangle\,.
\end{equation}
By using the fact that the two-point function of worldsheet fermions has the following property
\begin{equation}
\langle \bar{\psi}_2^{\mu}\bar{\psi}_1^{\nu} \rangle= \eta^{\mu\nu}\langle\bar{\psi}_2\bar{\psi}_1 \rangle\,
\end{equation}
we finally obtain that the contraction between $\partial X_1^\mu \partial X_2^\rho$ yields the contribution
\begin{align}
K_1=&- g_c^2 \epsilon_{1\mu\nu}\epsilon_{2\rho\sigma} p_{1\alpha}p_{2\gamma}\eta^{\mu\rho}\eta^{\alpha\sigma}\eta^{\nu\gamma} \langle \partial X_1\partial X_2\rangle \left(\langle\bar{\psi}_2\bar{\psi}_1\rangle\right)^2\\
=&-g_c^2 p_1^\sigma p_2^\nu\eta^{\mu\rho} \epsilon_{\mu\nu}\epsilon_{\rho\sigma}\langle \partial X_1\partial X_2\rangle \left(\langle\bar{\psi}_2\bar{\psi}_1\rangle\right)^2\,.
\end{align}
By renaming indices, we obtain
\begin{equation}
K_1=-g_c^2 \eta_{\mu\rho} p_{1\sigma}p_{2\nu}\epsilon^{\mu\nu}_1\epsilon_2^{\rho\sigma}\langle \partial X_1\partial X_2\rangle \left(\langle\bar{\psi}_2\bar{\psi}_1\rangle\right)^2\,.
\end{equation}

One can similarly sum over the other types of contractions. We report the result here
\begin{equation}\label{eqn:graviton correlators}
K=-g_c^2\eta_{\mu\rho} p_{1\sigma}p_{2\nu}\epsilon^{\mu\nu}_1\epsilon_2^{\rho\sigma} \left[ \langle \partial X_1\partial X_2\rangle (\bar{\psi}_2\bar{\psi}_1\rangle)^2+\langle \partial X_1\bar{\partial}X_2\rangle (\langle\psi_2\bar{\psi}_1\rangle)^2+c.c.\right]\,.
\end{equation}

\section{Doubling trick and two-point functions}\label{app:two point functions}
In this section, we summarize various two-point functions on the annulus, M{\"o}bius, and Klein bottle obtained by using the doubling trick. Following \cite{Antoniadis:1996vw,Polchinski:1998rq,Berg:2014ama}, we define the annulus $A,$ M{\"o}bius strip $\mathcal{M},$ and Klein bottle $\mathcal{K}$ by modding out the tori with modulus
\begin{equation}
\tau_A=it\,,\qquad \tau_{\mathcal{M}}=\frac{1}{2}+it\,,\qquad\tau_{\mathcal{K}}=2it\,
\end{equation}
by the involutions
\begin{equation}
I_A(z)=1-\bar{z}\,,\qquad I_{\mathcal{M}}(z)=1-\bar{z}\,,\qquad I_{\mathcal{K}}(z)=1-\bar{z}+\frac{\tau_\mathcal{K}}{2}\,.
\end{equation}
Note that our convention slightly differs from that of \cite{Antoniadis:1996vw,Berg:2014ama} as we are adopting the conventions of \cite{Polchinski:1998rq} for the worldsheet modulus. We will denote a Riemann surface of genus 1 with boundaries by $\sigma,$ and torus by $\mathcal{T}.$ 

To compute two-point functions on $\sigma,$ we will use the image charge method. Let us first start with bosonic correlators. On a torus, a bosonic correlator is 
\begin{equation}
\langle X(z_1)X(z_2)\rangle_{\mathcal{T}}=G_B(z_1,z_2;\tau)\,,
\end{equation}
where we define
\begin{equation}
 G_B(z_1,z_2;\tau)=-\frac{\alpha'}{2} \log \left|\frac{\vartheta_1(z_1-z_2|\tau)}{\vartheta_1'(0|\tau)}\right|^2+\frac{\pi \alpha'}{\tau_2}(\Im(z_1-z_2))^2\,.
\end{equation}
Note that the double derivative of the bosonic Green's function $G_B(z_1,z_2;\tau)$ satisfied the following identity
\begin{equation}
\partial_{z_1} \bar{\partial}_{z_1}G_B(z_1,z_2;\tau)= -\pi\alpha' \delta^2(z_1-z_2)+\frac{\pi\alpha'}{2\tau_2}\,,
\end{equation}
where we normalized the Dirac delta function as
\begin{equation}
\int_{\mathcal{T}} dz^2 \delta^2(z)=1\,,
\end{equation}
and the measure factor as
\begin{equation}
d^2z=2 dx dy\,.
\end{equation}
Note that the volume integral is therefore normalized as
\begin{equation}
\int_{\mathcal{T}} dz^2g_{z\bar{z}}=\tau_2\,.
\end{equation}
In this normalization, one can easily check 
\begin{equation}\label{eqn:vanishing int}
\int_{\mathcal{T}} dz^2 g_{z\bar{z}}\left( \partial_z\bar{\partial}_zG_B(z,0;\tau)\right)=0\,.
\end{equation}
It is important to note that for the physical correlator
\begin{equation}
\langle :\partial X(z_1): :\bar{\partial} X(z_2):\rangle\,,
\end{equation}
the Dirac delta term shall be omitted as explained in \cite{Antoniadis:1996vw}. This is because the operator $:\partial X(z):$ is holomorphically regulated.

Now let us compute the two-point function of bosons on $\sigma.$ Using the image charge method, we find
\begin{equation}
\langle X(z_1)X(z_2)\rangle_\sigma=\langle X(z_1)X(x_2)\rangle_\mathcal{T}+\langle X(z_1)X(I_\sigma(z_2))\rangle_\mathcal{T}\,.
\end{equation}
We compute the double derivatives of the Green's function, assuming the holomorphic regularization,
\begin{align}
\partial_{z_1}\partial_{z_2}\langle X(z_1)X(z_2)\rangle_\sigma=&\partial_{z_1}\partial_{z_2}G_B(z_1,z_2;\tau)+\frac{\pi\alpha'}{2\tau_2}\,,
\end{align}
\begin{equation}
\partial_{z_1}\bar{\partial}_{z_2}\langle X(z_1)X(z_2)\rangle_\sigma=\partial_{z_1}\bar{\partial}_{z_2}G_B(z_1,I_\sigma(z_2);\tau)-\frac{\pi\alpha'}{2\tau_2}\,.
\end{equation}
We note the following identities
\begin{equation}
\int_\sigma d^2z\left( f(z)+f(I_\sigma(z))\right)=\int_\mathcal{T} d^2z f(z)\,,
\end{equation}
and
\begin{equation}
\int_\sigma d^2z \left(\partial_zf(z)-\bar{\partial}_zf(I(z))\right)=\int_\mathcal{T}d^2z\partial_zf(z)\,.
\end{equation}
Using the fact that $G_B(z_1,z_2;\tau)$ is a symmetric function on torus, we find
\begin{equation}
\int_\sigma d^2z_1\int_\sigma d^2z_2\left(\partial_{z_1}\partial_{z_2}G_B(z_1,z_2;\tau)-\partial_{z_1}\bar{\partial}_{z_2}G_B(z_1,I_\sigma(z_2);\tau)+c.c\right)=0\,.
\end{equation}
Similarly, we compute
\begin{equation}
\int_\sigma g_{z_1\bar{z}_1}d^2z_1\int_\sigma g_{z_2\bar{z}_2}d^2 z_2\frac{\pi\alpha'}{2\tau_2}=\frac{\pi\alpha'}{8}\tau_2\,.
\end{equation}
As a result, we find
\begin{equation}\label{eqn:boson on sigma}
\int_\sigma g_{z_1\bar{z}_1}d^2z_1\int_\sigma g_{z_2\bar{z}_2}d^2 z_2 \left(\langle\partial X_1\partial X_2\rangle_\sigma-\langle \partial X_1\bar{\partial}X_2\rangle_\sigma+c.c\right)=\frac{\pi\alpha'}{2}\tau_2\,.
\end{equation}

Now, we summarize fermion two-point functions following the conventions of \cite{Berg:2014ama}. We will use a shorthand notation $s$ for the spin structure $(\alpha,\beta).$ Let us start with the correlation functions of fermions on the torus
\begin{equation}
\langle \psi (z_1)\psi(z_2)\rangle_{\mathcal{T}}^s=G_F(z_1,z_2;\tau,s)\,,\qquad \langle \bar{\psi}(\bar{z}_1)\bar{\psi}(\bar{z}_2)\rangle_\mathcal{T}^s=G_F(z_1,z_2;\tau,s)^*\,,
\end{equation}
\begin{equation}
\langle \psi(z_1)\bar{\psi}(\bar{z}_2)\rangle_\mathcal{T}^s=0\,,\qquad\langle\bar{\psi}(\bar{z}_1)\psi(z_2)\rangle_\mathcal{T}^s=0\,,
\end{equation}
where we define
\begin{equation}
G_F(z_1,z_2;\tau,s):= \frac{\vartheta_{\alpha,\beta}(z_1-z_2|\tau)\vartheta_{1,1}'(0|\tau)}{\vartheta_{\alpha,\beta}(0|\tau)\vartheta_{1,1}(z_1-z_2|\tau)}\,.
\end{equation}

As in the case of bosonic correlators, the fermionic two-point functions on $\sigma$ can be obtained by the image charge method. We summarize the fermionic two-point functions here
\begin{align}\label{eqn:fermion on sigma}
\langle \psi(z_1)\psi(z_2)\rangle_\sigma^s =&G_F(z_1,z_2;\tau,s)\,,\\
\langle\psi(z_1)\bar{\psi}(\bar{z}_2)\rangle_\sigma^s=&iG_F(z_1,I_\sigma(z_2);\tau,s)\,,\\
\langle\bar{\psi}(\bar{z}_1)\psi(z_2)\rangle_\sigma^s=&iG_F(I_\sigma(z_1),z_2;\tau,s)\,,\\
\langle\bar{\psi}(\bar{z}_1)\bar{\psi}(\bar{z}_2)\rangle_\sigma^s=&-G_F(I_\sigma(z_1),I_\sigma(z_2);\tau,s)\,.
\end{align}

\newpage
\bibliographystyle{JHEP}
\bibliography{refs}
\end{document}